\documentstyle[12pt,epsfig,psfig]{article}

\newcommand{\beq}{\begin{equation}}
\newcommand{\eeq}{\end{equation}}
\newcommand{\nn}{\nonumber}

\def\err{\end{array}}
\def\bea{\begin{eqnarray}}
\def\eea{\end{eqnarray}}
\def\nn{\noindent}
\def\inpb{\ifmmode {\rm pb}^{-1}\else ${\rm pb}^{-1}$\fi}
\def\infb{\ifmmode {\rm fb}^{-1}\else ${\rm fb}^{-1}$\fi}
\def\cc{\ifmmode k/\overline M_{Pl}\else $k/\overline M_{Pl}$\fi}
\newskip\zatskip \zatskip=0pt plus0pt minus0pt
\def\matth{\mathsurround=0pt}
\def\lsim{\mathrel{\mathpalette\atversim<}}
\def\gsim{\mathrel{\mathpalette\atversim>}}
\def\atversim#1#2{\lower0.7ex\vbox{\baselineskip\zatskip\lineskip\zatskip
  \lineskiplimit 0pt\ialign{$\matth#1\hfil##\hfil$\crcr#2\crcr\sim\crcr}}}

\begin{document}


\title{Models with Extra Dimensions \\
and Their Phenomenology}

\author{Yuri A. Kubyshin\\
Institute of Nuclear Physics, Moscow State University,\\
Moscow 119899, Russia\\
E-mail: kubyshin@theory.sinp.msu.ru}

\maketitle

\begin{abstract}
The Arkani-Hamed-Dimopoulos-Dvali and
the Randall-Sundrum models with extra spacelike dimensions, recently
proposed as a solution to the hierarchy
problem, are reviewed. We discuss their basic properties 
and phenomenological effects of particle interactions
at high energies, predicted within these models.
\end{abstract}


\section{Introduction} 
\label{intro}

Studies of field theoretical models in the space-time
with additional spatial dimensions were started by T. Kaluza and
O. Klein in their pioneering articles \cite{Kaluza}, \cite{Klein}
in the 20th. This gave the origin to a field theoretical approach 
to the description of particle interactions in a multidimensional 
space-time called the Kaluza-Klein (KK) approach.

Very detailed studies of various mathematical and physical aspects
of models within the KK approach were carried out in the literature,
see Refs. \cite{KK-review} - \cite{KaZu92} for reviews and references
therein. A strong motivation for the KK approach
comes from the string theory where the
multidimensionality of the space-time is required for the consistent
formulation.

Recently models with extra dimensions of a new type, namely the
Arkani-Hamed-Dimopoulos-Dvali
(ADD) and Randall-Sundrum (RS) models,  have been proposed
in Refs.~\cite{ADD1} - \cite{RS2}.
They were designed to provide a novel solution to the
hierarchy problem. Many essential features of these models
were inherited from the KK approach. Also many ideas and concepts,   
recently discovered in string/M-theories, have
been incorporated. The ADD and RS models are the subject of the
present contribution.

In the rest of this section we discuss main elements of the
"classical" KK theories. We also introduce the notion of localization of
fields on branes by considering a simple model and mention briefly
some recent results in string theories which serve as motivations
for the ADD and RS models. Sect. 2 and Sect. 3 are devoted
to the ADD model and the RS1 model respectively.
We explain the basics of the models and discuss possible effects, predicted
by them, which can be observed in high energy particle experiments.
We almost do not touch astrophysical effects and cosmological scenarios
within the ADD and RS models restricting ourselves to a very general
discussion and brief remarks. Sect. 4 contains a short summary of
results on the ADD and RS1 models.

In the present account we follow mainly the original papers on the subject
and some recent reviews, Refs.~\cite{Ant01} - \cite{Bes01}.

\subsection{Kaluza-Klein approach}
\label{intro:KK}

The KK approach is based on the hypothesis that
the space-time is a $(4+d)$-dimensional pseudo-Euclidean 
space 
\beq
E_{4+d} = M_{4} \times K_{d},   \label{E-space}
\eeq
where $M_{4}$ is the four-dimensional space-time
and $K_{d}$ is a $d$-dimensional compact space of
characteristic size (scale) $R$. The latter plays the role of 
the space of additional (spatial) dimensions of the space-time.
Let us denote local coordinates of $E_{4+d}$ as $\{\hat{x}^{M}\} = 
\{ x^{\mu}, y^{m}\}$, where $M=0,1,\ldots,3+d$, $\mu = 0,1,2,3$ and 
$m=1,2, \ldots, d$. In accordance with the direct product structure 
of the space-time, Eq. (\ref{E-space}), the metric is usually chosen to be
\[
ds^{2} = \hat{G}_{MN}(\hat{x}) d\hat{x}^{M} d\hat{x}^{N} =
g_{\mu \nu}(x) dx^{\mu} dx^{\nu} + \gamma_{mn}(x,y) dy^{m} dy^{n}.
\]

To illustrate main elements and ideas of the KK approach
which will be important for us later,
let us consider the case of $M_{4}=M^{4}$, the Minkowski space-time,
and a simple $(4+d)$-dimensional scalar model with the action
given by 
\beq
S = \int d^{4+d}\hat{x} \sqrt{-\hat{G}} \left[
-\frac{1}{2} \left( \partial_{M} \hat{\phi} \right)^{2} - 
\frac{m^{2}}{2} \hat{\phi}^{2} - \frac{g_{(4+d)}}{4!} \hat{\phi}^{4} \right], 
\label{sc-act}
\eeq
where $\hat{G}=\det (\hat{G}_{MN})$. 
To interprete the theory as an effective four-dimensional one
the multidimensional field $\hat{\phi}(x,y)$ is expanded in a Fourier
series
\beq
\hat{\phi} (x,y) = \sum_{n} \phi^{(n)}(x) Y_{n}(y), 
\label{KKexp}
\eeq
where $Y_{n}(y)$ are orthogonal normalized
eigenfunctions of the Laplace operator 
$\Delta_{K_{d}}$ on the internal space $K_{d}$, 
\beq
\Delta_{K_{d}} Y_{n}(y) = \frac{\lambda_{n}}{R^{2}} Y_{n}(y),  
\label{Y-ef}
\eeq
and $n$ is a (multi-) index labeling the eigenvalue
$\lambda_{n}$ of the eigenfunction $Y_{n}(y)$.

The case of the toroidal compactification of the
space of extra dimensions $K_{d}=T^{d}$, where $T^{d}$ denotes the 
$d$-dimensional torus with equal radii $R$, is particularly simple.
The multi-index $n=\{n_{1},n_{2}, \ldots, n_{d}\}$ 
with $n_{m}$ being integer numbers,
$-\infty \leq n_{m} \leq \infty$. The eigenfunctions
and eigenvalues in Eqs. (\ref{KKexp}), (\ref{Y-ef}) are given by
\bea
Y_{\{n_{1},n_{2}, \ldots n_{d} \}} & = & 
\frac{1}{\sqrt{V_{d}}} \exp \left\{\frac{i \sum_{m=1}^{d} n_{m}y^{m} }{R}
\right\}, 
\label{ef-torus} \\
\lambda_{\{n_{1},n_{2}, \ldots n_{d} \}} & = & n_{1}^{2} + n_{2}^{2} + 
\ldots n_{d}^{2},  \nonumber
\eea
where $V_{(d)} = (2\pi R)^{d}$ is the volume of the torus.

The coefficients $\phi^{(n)}(x)$ of the Fourier expansion (\ref{KKexp}) 
are called KK modes and play the role of fields of the effective
four-dimensional theory. Usually they include the zero-mode
$\phi^{(0)}(x)$, corresponding to $n=0$ and the eigenvalue
$\lambda_{0} = 0$. 
Substituting the KK mode expansion into action (\ref{sc-act}) and
integrating over the internal space $K_{d}$ one 
gets 
\bea
S & = & \int d^{4}x \sqrt{-g} \left\{  
-\frac{1}{2} \left( \partial_{\mu} \phi^{(0)} \right)^{2} - \frac{m^{2}}{2}
(\phi^{(0)})^{2} \right. \nonumber \\
& - & \left. \sum_{n \neq 0} \left[ \left(\partial_{\mu} \phi^{(n)} \right) 
\left(\partial^{\mu} \phi^{(n)} \right)^{*} + 
m_{n}^{2} \phi^{(n)}\phi^{(n)*} \right] \right. \nonumber \\
& - & \left. \frac{g_{(4)}}{4!} (\phi^{(0)})^{4} - 
\frac{g_{(4)}}{4} (\phi^{(0)})^{2} \sum_{n\neq 0} \phi^{(n)} \phi^{(n)*} 
- \ldots \nonumber \right\}, 
\eea
where the dots stand for the terms which do not contain the zero mode. 
The masses of the modes are given by 
\beq
m_{n}^{2} = m^{2} + \frac{\lambda_{n}}{R^{2}}. \label{mass}
\eeq
The coupling constant $g_{(4)}$ of the four-dimensional theory is related 
to the coupling constant $g_{(4+d)}$ of the initial multidimensional 
theory by the formula 
\beq
  g_{(4)} = \frac{g_{(4+d)}}{V_{(d)}},  \label{coupl}
\eeq
$V_{(d)}$ being the volume of the space of extra dimensions.  

Similar relations take place for other types of multidimensional theories. 
Consider the example of the Einstein $(4+d)$-dimensional gravity with 
the action
\[
S_{E} = \int d^{4+d}\hat{x} \sqrt{-\hat{G}} \frac{1}{16 \pi G_{N(4+d)}} 
{\cal R}^{(4+d)} [\hat{G}_{MN}], 
\]
where the scalar curvature ${\cal R}^{(4+d)} [\hat{G}_{MN}]$ is 
calculated using the metric $\hat{G}_{MN}$. 
Performing the mode expansion and integrating over $K_{d}$
one arrives at the four-dimensional action
\[
S_{E} = \int d^{4}x \sqrt{-g} \left\{ 
\frac{1}{16 \pi G_{N(4)}} {\cal R}^{(4)} [g^{(0)}_{MN}] + \ldots \right\}, 
\]
where the dots stand for the terms with non-zero modes. 
Similar to Eq. (\ref{coupl}), the relation between the
four-dimensional and $(4+d)$-dimensional gravitational
(Newton) constants is given by
\beq
G_{N(4)} = \frac{1}{V_{(d)}} G_{N(4+d)}, \label{G-rel}
\eeq
where, as before, $V_{(d)} \propto R^{d}$ is the volume
of the space of extra dimensions and $R$ is its size.
It is convenient to rewrite this relation in terms of the four-dimensional
Planck mass
$M_{Pl} = (G_{N(4)})^{-1/2} = 1.2 \cdot 10^{19}\; \mbox{GeV}$
and a fundamental mass scale of the $(4+d)$-dimensional theory 
$M \equiv (G_{N(4+d)})^{-\frac{1}{d+2}}$. We obtain 
\beq
M_{Pl}^{2} = V_{(d)} M^{d+2}. \label{M-rel}
\eeq
The latter formula is often referred to as the reduction formula. 

Initially the goal of the KK approach was to achieve a unification of
a few types of interactions in four dimensions within a unique interaction 
in the multidimensional space-time. A classical example is the
model proposed and studied in Ref. \cite{Kaluza}. It was shown there that 
the zero-mode sector of the five-dimensional Einstein gravity in
$E_{5} = M^{4} \times S^{1}$, where $M^{4}$ is the Minkowski space-time and
$S^{1}$ is the circle, is equivalent to the four-dimensional
theory which describes the Einstein gravity and electromagnetism. In this
model the electromagnetic potential appears from the $\hat{G}_{\mu 5}$  
components of the multidimensional metric. 

For multidimensional Yang-Mills theories with the space of extra
dimensions being a homogeneous space an elegant scheme of
dimensional reduction, called the coset space dimensional
reduction, was developed~\cite{CSDR} (see Refs.~\cite{KMRV89}, 
\cite{KaZu92} for reviews and Refs.~\cite{MaZu01} for recent studies).
An attractive feature of this scheme is that the effective four-dimensional 
gauge theory contains scalar fields. Their potential is of the 
Higgs type (i.e. leads to the spontaneous symmetry breaking) 
and appears in a purely geometrical way from the initial 
multidimensional Yang-Mills action. 

As we have seen, a characteristic feature of multidimensional
theories is the appearance of the infinite set of KK modes (called the
KK tower of modes). Correspondingly, a characteristic signature
of the existence
of extra dimensions would be detection of series of KK excitations
with a spectrum of the form (\ref{mass}).
So far no evidence of such excitations has been
observed in high energy experiments. The bound on the size $R$,
derived from the absence of signals of
KK excitations of the particles of the Standard Model (SM) in
the available experimental data, is
\[
  m_{1} \sim \frac{1}{R} \gsim 1  \; \; \mbox{TeV}   
\]
(see, for example, \cite{HEP-constr}).

\subsection{Localization of fields}
\label{localiz}

\begin{figure}[ht]
\begin{center}
\epsfig{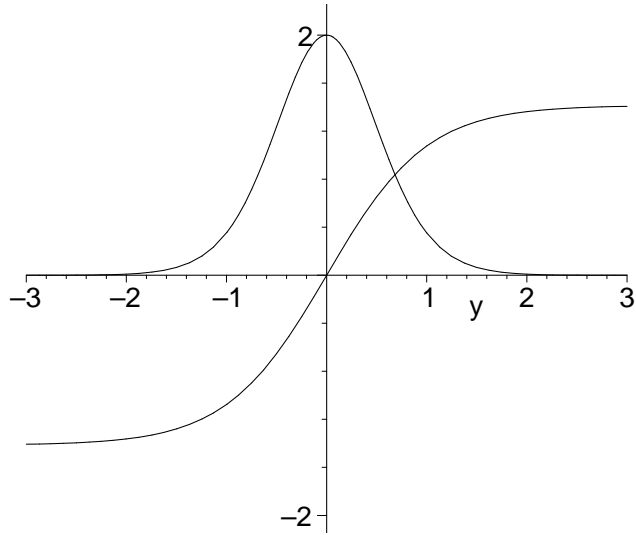}
\end{center}
\caption{The kink solution and its energy density in the scalar model 
with potential (\ref{kink-V}).}
\label{fig:kink}
\end{figure}

A new ingredient which turned out to be essential for the recent 
developments was elaborated by V. Rubakov and M. Shaposhnikov 
in article \cite{RubSh} (see also Ref. \cite{Akama}). 
This is the mechanism of localization of fields on branes. 
The authors considered the theory of the scalar field $\Phi (x,y)$
with a quartic potential in the five-dimensional space-time
$M^{4} \times R^{1}$, i.e. with the infinite fifth dimension.
Here the coordinate $\hat{x}^{4} \equiv y^{1}$ is denoted by $y$.
The potential was chosen to be
\beq
V(\Phi) = \frac{\lambda}{4} \left( \Phi^{2} - \frac{m^{2}}{\lambda} 
\right)^{2} =  -\frac{m^{2}}{2} \Phi^{2} + \frac{\lambda}{4} \Phi^{4} 
+ \frac{m^{4}}{4\lambda}.
\label{kink-V}
\eeq
It is easy to check that there exists a kink solution
which depends on the fifth coordinate only. The solution
is given by
\[
\Phi_{cl} (y) = \frac{m}{\sqrt{\lambda}} \tanh \frac{m(y-y_{0})}{\sqrt{2}}.
\]
The kink solution centered at $y=y_{0}=0$ is shown in
Fig. \ref{fig:kink}. The energy density of this configuration is
localized in the vicinity of the hyperplane $y=y_{0}$ 
within a region of thickness $\sim 1/m$.
The spectrum of quantum fluctuations (KK modes) around the kink solution
includes a zero mode (corresponding to the translational symmetry
of the theory), one massive mode and a continuum of states. 
For low enough energies only the discrete modes are excited, 
and effectively the theory describes fields
moving inside the potential well along the plane $y_{0}=0$.
The model provides an example of dynamical
localization of fields on the hyperplane which plays the role of our
three-dimensional space embedded into the four-dimensional space.
This hyperplane is referred to as a wall or 3-brane.
If the energy is high enough modes of the continuous spectrum are excited, 
and a manifestation of this may be effects with particles escaping into 
the fifth dimension. 

In a similar way fermions coupled to the scalar field can be localized 
on the wall. Localization of vector fields was discussed in Ref. \cite{DvSh}.
A field theoretical and string theory realizations of the localization 
mechanism are discussed in Refs. \cite{ADD1}, \cite{AAHDD98}. 
Within the string theory framework localization of vector fields on branes 
is automatic (see \cite{Polch}).  

\subsection{String/M-theory motivations}
\label{string}

In recent years there were a number of developments in string theories which
motivated the models with additional dimensions we are going
to discuss in Sect.~2 and Sect.~3.

An important case is when the space of extra dimensions is an orbifold. In
this paper we will be dealing only with the orbifold $S^{1}/Z_{2}$ which
is defined in the following way. Let us consider the circle $S^{1}$ of the
radius $R$ and 
denote its points by $y$. The orbifold is constructed by identifying
the points which are related by the $Z_{2}$-transformations $y \rightarrow
(-y)$. We will write this orbifold $Z_{2}$-identification as 
$y \cong (-y)$. In addition we have the usual identification of points of
$S^{1}$ due to periodicity: $y \cong (y + 2\pi R)$. 
The points $y=0$ and $y=\pi R$ are fixed points of the
$Z_{2}$-identification. 3-branes
\footnote{A $p$-brane is understood as a hypersurface with $p$ spacelike
dimensions embedded into a larger space-time.}, or
three-dimensional hyperplanes, playing the role of
our three-dimensional space, can be located at these fixed points.  
  
In the case of the orbifold compactification all
multidimensional fields either even or odd. 
In the former case they satisfy the condition $\Phi (x,y) = \Phi (x,-y)$,
whereas in the latter one they satisfy $\Phi (x,y) = - \Phi (x,-y)$. Their
mode decompositions are of the form
\bea
\Phi (x,y) & = & \sum_{n=0}^{\infty} \Phi^{(n)}(x) \cos \frac{ny}{R} 
\; \; \mbox{if} \; \; \Phi (x,y) \; \; \mbox{is even},  \nonumber \\
\Phi (x,y) & = & \sum_{n=1}^{\infty} \Phi^{(n)}(x) \sin \frac{ny}{R} 
\; \; \mbox{if} \; \; \Phi (x,y) \; \; \mbox{is odd}.  
\nonumber 
\eea  
Note that only even fields contain zero modes. Correspondingly, only
even fields are non-vanishing on the branes at $y=0$ and $y=\pi R$. 

One important feature of string theories is that there exist  
$p$-brane configurations which confine gauge and other degrees 
of freedom naturally (see \cite{Polch}, \cite{string}). Another one is that 
there are consistent compactifications of the 11D limit of M-theory 
\cite{HoWi96a} - \cite{Wi96}, 
regarded as a theory of everything, down to 
$(3+1)$ dimensions. The path of compactification is not unique. 

In Refs. \cite{HoWi96a},  \cite{HoWi96b} 
it was shown that the non-perturbative regime 
of the $E_{8} \times E_{8}$ heterotic string can be identified as the 11D 
limit of M-theory with one dimension being compactified to the 
$S^{1}/Z_{2}$ orbifold and with a set of $E_{8}$ gauge fields at each 10D 
fixed point of the orbifold. By compactifying this theory on the 
Calabi-Yau manifold one arrives at a theory which, 
for certain range of energies, behaves like a 5D 
${\cal N}=1$ supersymmetric (SUSY) model with one dimension compactified on 
$S^{1}/Z_{2}$ \cite{Wi96}, \cite{BaDi96}. The effective action 
of this theory was derived in \cite{LOSW99}. 

\begin{figure}[tp]
\epsfbox{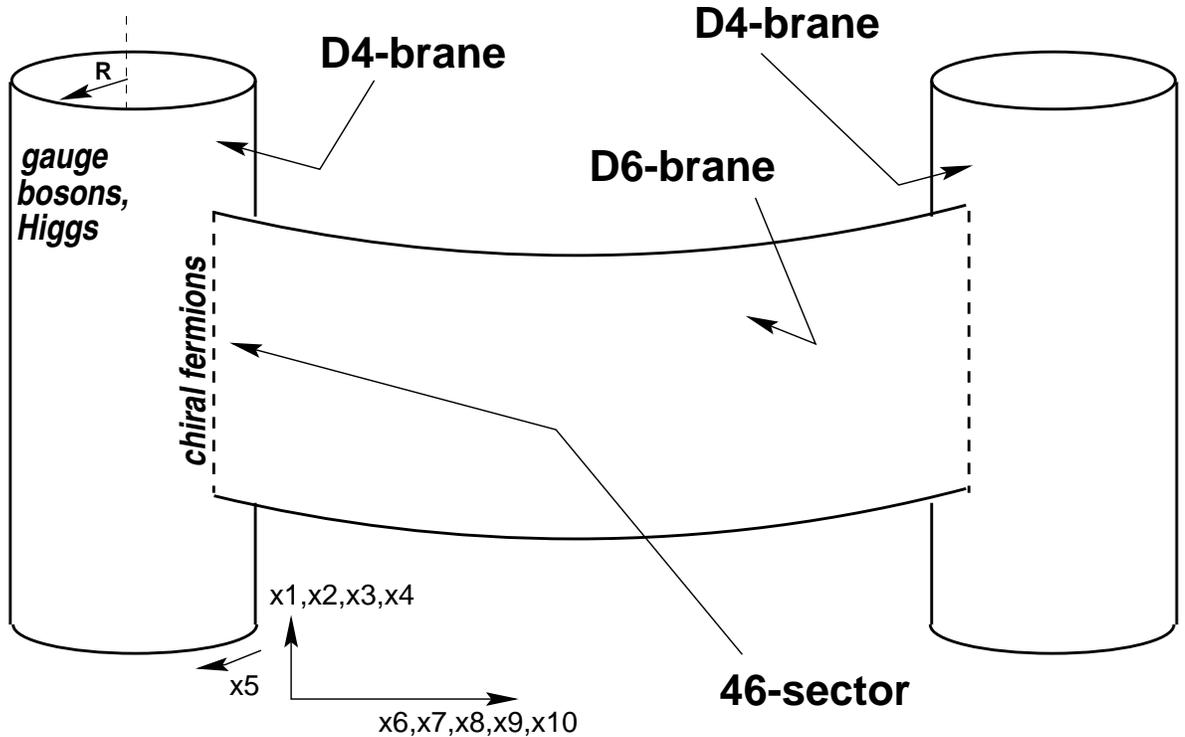}
\caption{Schematical picture of the space-time in the 
Type I$^{\prime}$ string theory, taken from Ref. \cite{DDG}.}
\label{fig:brane-bulk}
\end{figure}

The main features of string motivated scenarios, considered in the 
literature, are the following.
The ends of open strings are restricted to $D4$-branes 
in the 11-dimensional space-time. Excitations of open strings 
include gauge bosons, scalar fields and fermions. A  
low energy effective theory of these excitations is supposed to contain 
the SM, hence the SM fields are localized on the  
$D4$-brane. Closed strings propagate in the bulk, i.e. in the whole 
space-time. Excitations of closed strings include the graviton, 
therefore gravity propagates in the bulk. A schematical picture 
of the space-time in the Type I$^{\prime}$ string theory, as viewed in 
Ref. \cite{DDG}, is presented in Fig. \ref{fig:brane-bulk}. The $D4$-brane 
contains three non-compact spacelike dimensions, corresponding to 
our usual space, and one dimension compactified to the circle 
$S^{1}$ of the radius $r$. As far as the rest of the dimensions
are concerned, $5$ spatial dimensions or a part of them   
may form a compact space of the characteristic size $R$. 
A systematic effective field theory model of a 3-brane 
Universe was considered in Ref.~\cite{Sund98}. In Ref.~\cite{ANO00} 
the creation of brane-worlds within the minisuperspace 
restriction of the canonical Wheeler-DeWitt formalism was discussed.  

The assumption $R \gg r$ does not seem to be impossible. 
It is motivated by the fact that in string theory 
the tree-level formula  
\[
  M_{str} = M_{Pl} \frac{\sqrt{k \alpha_{GUT}}}{2}
\]
between the Planck mass $M_{Pl}$ and the 
string scale $M_{str}$ receives large corrections. Therefore 
the relations $M_{str} \ll M_{Pl}$ and 
$R \sim M^{-1}_{str} \gg M^{-1}_{Pl} \sim r$ are 
not inconsistent with the string theory 
\cite{HoWi96b}, \cite{Wi96}, \cite{Lyk96}. 
With this possibility in mind there were 
proposals of scenarios where part of the compact dimensions had  
relatively large size. For example, in a scenario shown in Fig. 
\ref{fig:brane-bulk} a physically interesting case corresponds to 
$R^{-1} \sim 10$MeV, $r^{-1} \sim 1$TeV. 

Additional motivations of scenarios with large extra dimensions 
are the following:

1) possibility of unification of strong and electroweak interactions
at a lower scale $M'_{GUT} \sim 10$TeV due to the power law running of 
couplings in multidimensional models \cite{DDG98} 
(see also \cite{DDG}); 

2) existence of novel mechanisms of SUSY breaking 
\cite{SUSY1} - \cite{GhPo2} and 
electroweak symmetry breaking \cite{EWSB}. 
For example, one of them is to break SUSY on the hidden brane, 
then the SUSY breaking is communicated to the SM brane via  
fields in the bulk (see, for example, \cite{SUSY1}, \cite{KaTa01}).  

We would like to mention that studies of the 
power law running of couplings in the minimal supersymmetric standard 
model within the exact renormalization group approach were carried out 
in Ref.~\cite{KKMZ98}, various aspects of the running of coupling 
constants in KK theories were addressed, for example, 
in Refs.~\cite{KK-run}.  

\begin{figure}[ht]
\begin{center}
\epsfig{file=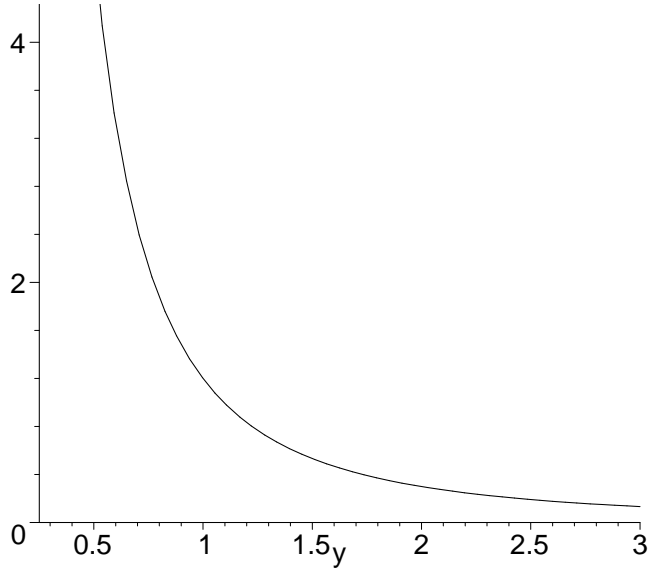,%
bbllx=120pt,bblly=250pt,%
bburx=490pt,bbury=480pt,%
clip=%
}
\end{center}
\caption{Deviation of the cross section in a KK model as a function 
of the inverse size $R^{-1}$ of extra dimensions.}
\label{fig:Delta-R}
\end{figure}

\begin{figure}[ht]
\begin{center}
\epsfig{file=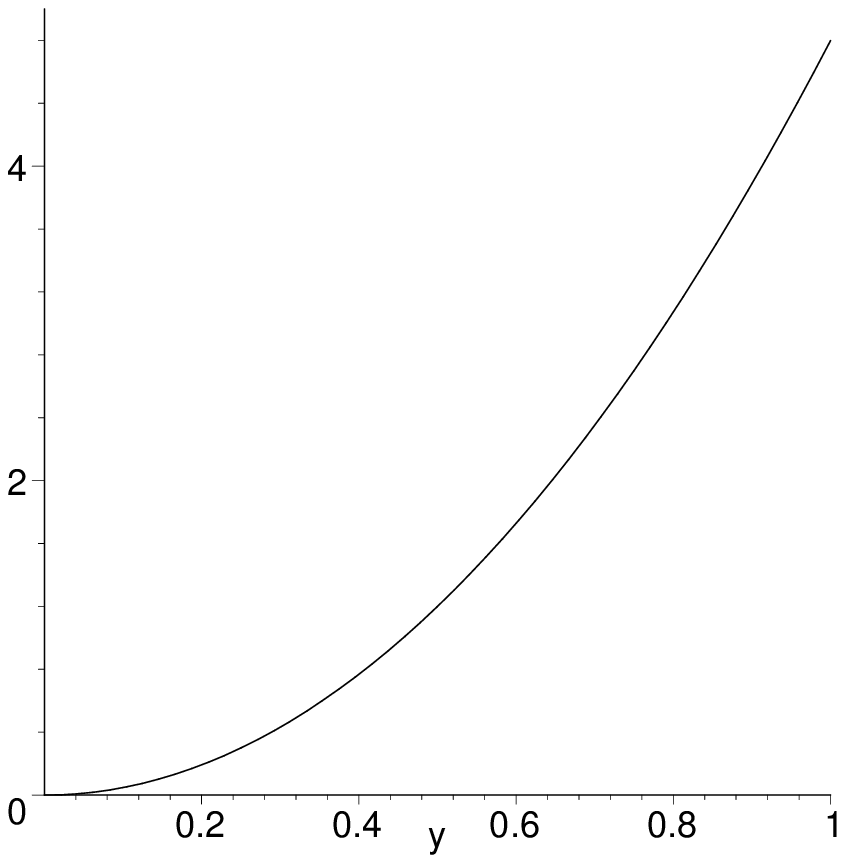,%
bbllx=120pt,bblly=250pt,%
bburx=490pt,bbury=480pt,%
clip=%
}
\end{center}
\caption{Deviation of the cross section in a KK model as a function 
of $\sqrt{s}/(2m_{1})$, where $\sqrt{s}$ is the centre-of-mass 
energy of colliding particles and $m_{1}$ is the mass of the first 
non-zero KK mode.}
\label{fig:Delta-s}
\end{figure}

In fact, extensive studies of 
physical effects and phenomenological predictions  
of Kaluza-Klein theories with the size of extra dimensions 
$R \sim 1 \; \mbox{TeV}^{-1}$ have been started almost a decade ago
(see \cite{Ant90} - \cite{DKPC}, \cite{PoQui98}  
and references therein). 
In particular, it was shown that in such models the 
cross section $\sigma$ 
of a given process deviates from the cross section $\sigma^{(SM)}$ 
of the SM in four dimensions even for 
energies below the threshold of the creation of the first non-zero mode 
in the KK tower. This effect is due to contributions of the 
processes of the exchange via virtual KK  
excitations. A characteristic dependence of   
$\Delta = (\sigma - \sigma^{(SM)})/\sigma^{(SM)}$ on the inverse 
size of extra dimensions $R^{-1}$ and on the centre-of-mass energy 
$\sqrt{s}$ of colliding particles is shown in Figs. \ref{fig:Delta-R}, 
\ref{fig:Delta-s}.


\vspace{0.5cm}

Above we explained basics of the traditional KK approach, which 
will be essential to us. We also outlined some recent ideas, which lie
in the basis 
of the new type KK models. Whereas the main goal of the "classical" 
KK theories was unification of various types of interactions within 
a more fundamental interaction in the multidimensional
space-time, the aim of the new models with extra dimensions is 
to solve the {\it hierarchy problem}. The essence of this problem 
is the huge difference between the fundamental Planck scale $M_{Pl}$,
which is also the scale of the gravitational interaction, 
and the electroweak scale $M_{EW} \sim 1$TeV. 
Namely, $M_{EW}/M_{Pl} \sim 10^{-16}$. 

\section{Arkani-Hamed - Dimopoulos - Dvali Model}
\label{ADD:model}

\subsection{Main features of the model}

We consider first the ADD model proposed and studied by N. Arkani-Hamed, 
S. Dimopoulos and G. Dvali in Refs. \cite{ADD1}, \cite{ADD2}. 
The model includes the SM localized on a 3-brane 
embedded into the $(4+d)$-dimensional space-time with compact 
extra dimensions. The gravitational field is the only field which 
propagates in the bulk. This construction is usually viewed as a part of a 
more general string/M-theory with, possibly, other branes
and more complicated geometry of the space of extra dimensions. 
Reduction formula (\ref{M-rel}) is valid in this case too.
Since the volume of the space of extra dimensions $V_{(d)} \sim R^{d}$, 
where $R$ is its characteristic size, the reduction formula 
can be written as
\beq
M_{Pl}^{2} \sim R^{d} M^{2+d}  \label{M-rel1}
\eeq
(we assume for simplicity that all $d$ extra dimensions are of the same
size). 
The hierarchy problem is avoided simply by removing the hierarchy, namely 
by taking the fundamental mass scale $M$ of the multidimensional 
theory to be $M \sim 1$TeV. In this way $M$ becomes the only 
fundamental scale both for gravity and the electroweak interactions. 
At distances $r \lsim R$ the gravity is essentially $(4+d)$-dimensional. 
Using the value of the Planck mass Eq. (\ref{M-rel1}) can be rewritten as 
\[
R \sim \frac{1}{M} \left( \frac{M_{Pl}}{M} \right)^{2/d} \sim 
10^{\frac{30}{d}-17} \; \mbox{cm},   
\]
or
\[
R^{-1} \sim M \left( \frac{M}{M_{Pl}} \right)^{2/d} \sim 
10^{-\frac{30}{d}+3} \; \mbox{GeV}   
\]
\cite{ADD2}.

Let us analyze various cases. In the case $d=1$ it follows from the 
formulas above that $R \sim 10^{13} \mbox{cm}$, i.e. the size of 
extra dimensions is of the order of the solar distance. This case is 
obviously excluded. For $d \geq 2$ we obtain:  

\begin{tabular}{lll}
    &   & \\
for $d=2$ & $R \sim 0.1 \; \mbox{mm}$, & 
     $R^{-1} \sim 10^{-3} \; \mbox{eV}$ \\
for $d=3$ \ \ & $R \sim 10^{-7} \; \mbox{cm}$, & 
     $R^{-1} \sim 100 \; \mbox{eV}$ \\
  $\ldots$  \ \ &   $\ldots$  \ \ &   $\ldots$  \ \   \\     
for $d=6$ \ \ & $R \sim 10^{-12} \; \mbox{cm}$, & 
     $R^{-1} \sim 10 \; \mbox{MeV}$ \\
     &  &  \\
\end{tabular}

Such size of extra dimensions is already acceptable because no deviations 
from the Newtonian gravity have been observed for distances 
$r \lsim 1$mm so far (see, for example, \cite{Hoyle}). 
On the other hand, as it was already mentioned in the previous section, 
the SM gauge forces have been accurately measured already 
at the scale $\sim 100$GeV. Hence, for the model to be consistent 
fields of the SM must be localized on the 3-brane. Therefore, only gravity 
propagates in the bulk. 

Let us assume for simplicity that the space of extra dimensions is the 
$d$-dimensional torus. To analyze the field content of the effective 
(dimensionally reduced) four-dimensional model 
we first introduce the field $\hat{h}_{MN}(x,y)$, 
describing the linear deviation of the metric around the $(4+d)$-dimensional 
Minkowski background $\eta_{MN}$, by 
\beq
\hat{G}_{MN} (x,y) = \eta_{MN} + \frac{2}{M^{1+d/2}} \hat{h}_{MN}(x,y) 
\label{ADD:Gh}
\eeq 
and then perform the KK mode expansion 
\beq
\hat{h}_{MN} (x,y) = \sum_{n} h^{(n)}_{MN} (x) \frac{1}{\sqrt{V_{(d)}}} 
e^{-i\frac{n_{m}y^{m}}{R}}  \label{ADD-exp}
\eeq
(cf. (\ref{KKexp}), (\ref{ef-torus})), where $V_{(d)}$ is the volume of the 
space of extra dimensions. The masses of the KK modes are given by 
\beq
m_{n} = \frac{1}{R} \sqrt{n_{1}^{2}+n_{2}^{2}+ \ldots + n_{d}^{2}} 
\equiv \frac{|n|}{R},  
\label{ADD-mass}
\eeq
so that the mass splitting of the spectrum is  
\beq
\Delta m \propto 1/R.    \label{ADD-split}
\eeq

The Newton potential between two test masses $m_{1}$ and $m_{2}$, 
separated by a distance $r$, is equal to 
\[
V(r) = G_{N(4)} m_{1}m_{2} \sum_{n} \frac{1}{r} e^{-m_{n}r} 
= G_{N(4)} m_{1}m_{2} \left( \frac{1}{r} + 
   \sum_{n \neq 0} \frac{1}{r} e^{-|n|r/R}  \right). 
\]
The first term in the last bracket is the contribution of the 
usual massless graviton (zero mode). The second term is the sum 
of the Yukawa potentials due to contributions of the massive gravitons. 
For the size $R$ large enough (i.e. for the spacing between the modes small 
enough) this sum can be substituted by the integral and we get \cite{ADD2}
\bea
V(r) & = & G_{N(4)} \frac{m_{1}m_{2}}{r} \left[
1 + S_{d-1} \int_{1/R}^{\infty}e^{-mr} m^{d-1} dm \right] \nonumber \\
     & = & G_{N(4)} \frac{m_{1}m_{2}}{r} \left[
1 + S_{d-1} \left( \frac{R}{r} \right)^{d} \int_{r/R}^{\infty} 
e^{-z} z^{d-1} dz \right], \label{ADD-Vprim} 
 \label{ADD-V1}
\eea 
where $S_{d-1}$ is the area of the $(d-1)$-dimensional sphere $S^{d-1}$ 
of the unit radius. 
It is easy to see that when the distance between the test masses $r \gg R$ 
the potential $V(r)$ is the usual Newton potential in four dimensions, 
\[
V(r) \approx G_{N(4)} \frac{m_{1} m_{2}}{r}. 
\]
At short distances, when $r \ll R$, the second term in Eq. (\ref{ADD-Vprim})
dominates so that 
\[
V(r) \approx G_{N(4)} \frac{m_{1} m_{2}}{r} S_{d-1} 
     \left( \frac{R}{r} \right)^{d} \Gamma (d) = 
     G_{N(4+d)} \frac{m_{1} m_{2}}{r^{d+1}} S_{d-1} \Gamma (d),   
\]
i.e. the $(4+d)$-dimensional gravity law is restored. Here we used 
relation (\ref{G-rel}) which is essentially the reduction formula, Eq. 
(\ref{M-rel1}). 

Using simple considerations, it was shown in Ref. \cite{ADD2} that 
the model, described here, is pretty consistent. In particular, for $d=2$ the 
deviation $\Delta E_{grav}$ of the gravitational energy of a simple 
system from its four-dimensional gravitational 
energy $E_{grav}$ at a distance $r$ was estimated to be 
\[
\left. \frac{\Delta E_{grav}}{E_{grav}} \right|_{r} \sim 
\left( \frac{R}{r} \right)^{2} \sim 
\left( \frac{1 \; \mbox{mm}}{r} \right)^{2}.
\]
Though at atomic distances gravity becomes multidimensional it is still 
much weaker than the electromagnetic forces and still need not to be 
taken into account. For example, for $d=2$ the ratio of the gravitational 
force between an electron and a positron 
to the electromagnetic attractive force between them at distances 
$r \sim 10^{-8} \; \mbox{cm}$ was estimated to be
\[
\frac{F_{grav}}{F_{em}} \sim 10^{-25}. 
\]

Let us discuss now the interaction of the KK modes $h^{(n)}_{MN}(x)$ with 
fields on the brane. It is determined by the universal minimal 
coupling of the $(4+d)$-dimensional theory:
\[
S_{int} = \int d^{4+d}\hat{x} \sqrt{-\hat{G}} \hat{T}_{MN} 
\hat{h}^{MN} (x,y),  
\]
where the energy-momentum tensor of matter on the brane localized at 
$y=0$ is of the form  
\[
\hat{T}_{MN}(x,y) = \delta_{M}^{\mu} \delta_{N}^{\nu} T_{\mu \nu}(x) 
\delta^{(d)}(y). 
\]
Using the reduction formula, Eq. (\ref{M-rel1}), and KK expansion 
(\ref{ADD-exp}) we obtain that 
\bea
S_{int} & = & \int d^{4}x T_{\mu \nu} \sum_{n} 
\frac{1}{M^{1+d/2} \sqrt{V_{(d)}}} h^{(n)\mu \nu} (x)   \nonumber \\
& \sim & \sum_{n} \int d^{4}x \frac{1}{M_{Pl}} 
T^{\mu \nu}(x) h^{(n)}_{\mu \nu}. \label{ADD-int1}
\eea

The degrees of freedom of the four-dimensional theory, which emerge from the 
multidimensional metric, are described in Refs. \cite{GRW}, \cite{Hew}.  
They include: (1) the massless graviton and the massive KK gravitons 
(spin-2 fields) with masses given by Eq.~(\ref{ADD-mass}),  
we will denote the gravitons by $G^{(n)}$;  
(2) $(d-1)$ KK towers of spin-1 fields which do not couple to $T_{\mu \nu}$; 
(3) $(d^{2}-d-2)/2$ KK towers of real scalar fields (for $d \geq 2$), they 
do not couple to $T_{\mu \nu}$ either; 
(4) a KK tower of scalar fields coupled to the trace of the 
energy-momentum tensor $T_{\mu}^{\mu}$, its zero mode is called radion and 
describes fluctuations of the volume of extra dimensions. 
We would like to note that for the model to be viable a mechanism 
of the radion stabilization need to be included. With this modification 
the radion becomes massive. 

As in the "classical" KK approach, there are two equivalent pictures which 
can be used for 
the analysis of the model at energies below $M$. One can consider 
the $(4+d)$-dimensional theory with the $(4+d)$-dimensional massless 
graviton $\hat{h}_{MN}(x,y)$ interacting with the SM fields with 
couplings $\sim 1/M^{1+d/2}$. Equivalently, one can consider the effective 
four-dimensional theory with the field content described above. 
In the latter picture the coupling of each individual 
graviton (both massless and massive) to the SM fields is small, namely 
$\sim 1/M_{Pl}$. However, the smallness of the coupling constant is 
compensated by the high multiplicity of states with the same mass. 
Indeed, the number $d{\cal N}(|n|)$ of modes with the modulus $|n|$ of the 
quantum number being in the interval $(|n|,|n|+d|n|)$ is equal to 
\beq
d{\cal N}(|n|) = S_{d-1} |n|^{d-1}d|n| = S_{d-1} R^{d} m^{d-1}dm \sim  
S_{d-1} \frac{M_{Pl}}{M^{d+2}} m^{d-1}dm,   \label{ADD-dN}
\eeq
where we used the mass formula $m=|n|/R$ and the reduction formula, Eq. 
(\ref{M-rel1}). The number of KK gravitons $G^{(n)}$ with masses 
$m_{n} \leq E < M$ is equal to 
\[
{\cal N}(E) \sim \int_{0}^{ER} d{\cal N}(|n|) \sim 
S_{d-1} \frac{M^{2}_{Pl}}{M^{d+2}} \int_{0}^{E} m^{d-1}dm = 
\frac{S_{d-1}}{d} \frac{M^{2}_{Pl}}{M^{d+2}} E^{d} \sim R^{d} E^{d}. 
\]
Here we used integration instead of summation. As it was already 
mentioned above, this is justified because of smallness of 
the mass splitting (see Eq.~(\ref{ADD-split})). 
We see that for $E \gg R^{-1}$ the multiplicity of states which can be 
produced is large. According to Eq. (\ref{ADD-int1}) the amplitude of 
emission of the mode $n$ is ${\cal A} \sim 1/M_{Pl}$, and correspondingly 
its rate is $|{\cal A}|^{2} \sim 1/M_{Pl}^{2}$. The total combined rate 
of emission of the KK gravitons with masses $m_{n} \leq E$ is 
\beq
\sim \frac{1}{M_{Pl}^{2}} {\cal N}(E) \sim \frac{E^{d}}{M^{d+2}}. 
\label{ADD-rate}
\eeq
We see that there is a considerable enhancement of the effective coupling 
due to the large phase space of KK modes, or, in another way of saying, 
due to the large volume of the space of extra dimensions. Because 
of this enhancement the cross-sections of processes involving 
the production of KK gravitons may turn out to 
be quite noticeable at the future colliders. 

The ADD model is often considered in the approximation when 
the wall is infinitely heavy and is described by a hyperplane of 
zero thickness fixed at, say, $y=0$, i.e. the translational 
invariance along the internal space is broken. 
In this case the discrete momentum $n_{m}$ along the extra dimensions 
is not conserved in the interactions on the brane. 
The energy, however, is conserved. 
In the bulk the discrete momentum $n_{m}$ is conserved. So, for example, 
the decay $G^{(n)} \rightarrow G^{(k)}G^{(l)}$ of the massive KK graviton 
into lighter KK modes $G^{(m)}$, $G^{(k)}$ is possible only if 
$n_{m}=k_{m}+l_{m}$, $m=1,2, \ldots , d$. 

Let us now analyze decays of the massive gravitons into the SM 
particles, for example, $G^{(n)} \rightarrow \gamma \gamma$. Such processes 
are very different from decays of four-dimensional particles on the 
brane. Intuitively it is clear that the gravitons escape into the 
bulk with a low probability of returning to interact with the SM 
fields on the brane. As it was estimated in Ref. \cite{ADD2} the 
total width of such decay is equal to 
$\Gamma_{n} = P \times \Gamma_{n}'$. Here $P$ is the 
probability of the graviton to happen to be near the brane. It can be 
estimated as the ratio of the $d$th power of the Compton wavelength to 
the volume $V_{(d)}$ of the space of extra dimensions, so that 
\[
  P \sim \frac{(1/m_{n})^{d}}{V_{(d)}} \sim \frac{1}{R^{d} m_{n}^{d}}. 
\]
$\Gamma_{n}'$ is the width of the decay of the graviton into the SM 
particles, $\Gamma_{n}' \sim m_{n}^{d+3}/M^{d+2}$. Combining these 
expressions and using the reduction formula we obtain that 
\beq
\Gamma_{n} \sim \frac{m_{n}^{3}}{M_{Pl}^{2}}. \label{ADD-decay} 
\eeq
This formula can be also understood in a different way. 
The factor $1/M_{Pl}^{2}$ reflects the fact that each mode 
interacts with the coupling suppressed by the Planck mass. Then the factor 
$m_{n}^{3}$ in Eq. (\ref{ADD-decay}) can be restored by dimensional 
considerations. The lifetime of the mode $n$ is equal to 
\beq
\tau_{n} = \frac{1}{\Gamma_{n}} = \frac{1}{M_{Pl}}  
\left( \frac{M_{Pl}}{m_{n}} \right)^{3} = 
10^{31} \; \mbox{yr} \; \left( \frac{1 \; \mbox{eV}}{m_{n}} \right)^{3}.
\label{ADD-tau}
\eeq

We see that the KK gravitons behave like massive, almost stable 
non-interacting spin-2 particles. A possible collider signature is 
imbalance in final state momenta and missing mass. Since the 
mass splitting $\Delta m \sim 1/R$ the inclusive cross-section reflects 
an almost continuous distribution in mass. This characteristic feature 
of the ADD model may enable to distinguish its predictions from other 
new-physics effects.  

\subsection{HEP phenomenology} 

The Feynman rules for diagrams involving gravitons in the ADD model were 
derived in Refs. \cite{GRW}, \cite{HLZ98}.  The high energy 
effects predicted by the model were studied in \cite{GRW}, 
\cite{MiPePe99} - \cite{ChKe}, and many others.  

There are two types of processes at high energies 
in which effects due to KK modes of the graviton can be observed at 
running or planned experiments. These are the graviton emission 
and virtual graviton exchange processes. 

We start with the graviton emission, i.e. the reactions 
where the KK gravitons are created as final state particles. 
These particles escape from the detector, so that a characteristic 
signature of such processes is missing energy. As we explained 
above, though the rate of production of each individual mode 
is suppressed by the Planck mass, due to the high multiplicity 
of KK states, Eq. (\ref{ADD-dN}), the magnitude of the 
total rate of production is determined by the TeV scale (see Eq. 
(\ref{ADD-rate})). Taking Eq. (\ref{ADD-dN}) into account, 
the relevant differential cross section \cite{GRW}
\beq
\frac{d^{2}\sigma}{dt dm} \sim  S_{d-1} \frac{M^{2}_{Pl}}{M^{d+2}} 
m^{d-1} \frac{d \sigma_{m}}{dt} \sim \frac{1}{M^{d+2}},   
    \label{ADD-sigma}
\eeq
where $d \sigma_{m}/dt $ is the differential cross section of the 
production of a single KK mode with mass $m$. 

\begin{figure}[tp]
\epsfxsize=0.9\hsize
\epsfbox{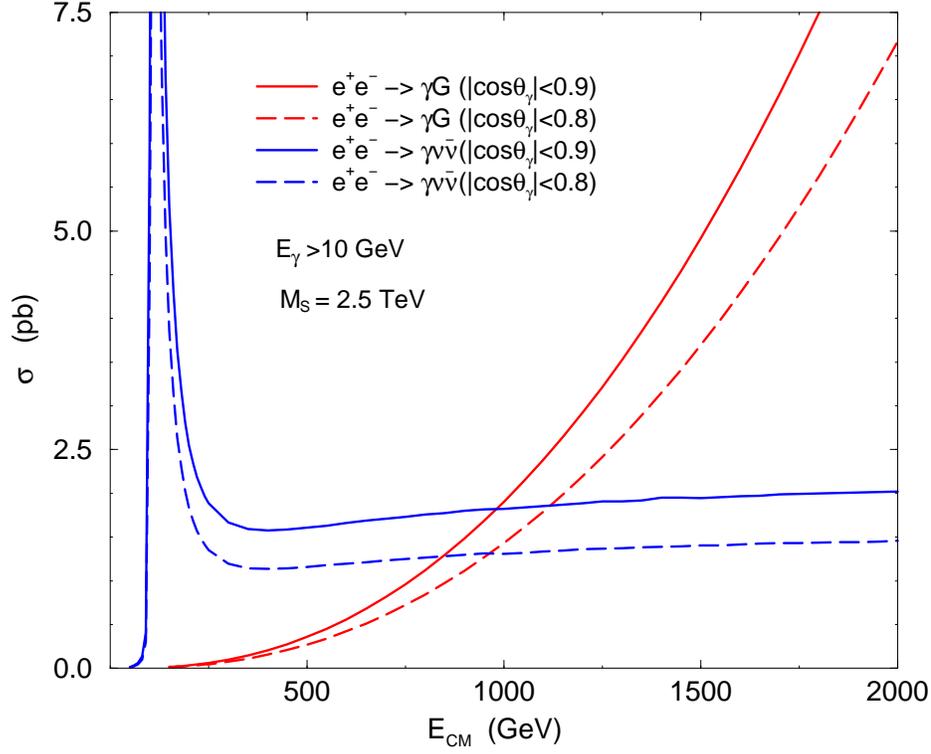}
\caption{The total cross sections for 
$e^{+}e^{-} \rightarrow \gamma \nu_{i} \bar{\nu}_{i}$ ($i=e,\mu,\tau$) 
(curves with the peak) and 
$e^{+}e^{-} \rightarrow \gamma G$ (faster growing curves) 
with cuts as shown for $d=2$ and $M = 2.5$ TeV \cite{ChKe}. 
On the plot $M$ is denoted as $M_{S}$. The 
peak corresponds to the threshold of the $Z$-boson. }
\label{fig:ADD-ee}
\end{figure}

\begin{figure}[tp]
\epsfxsize=0.9\hsize
\epsfbox{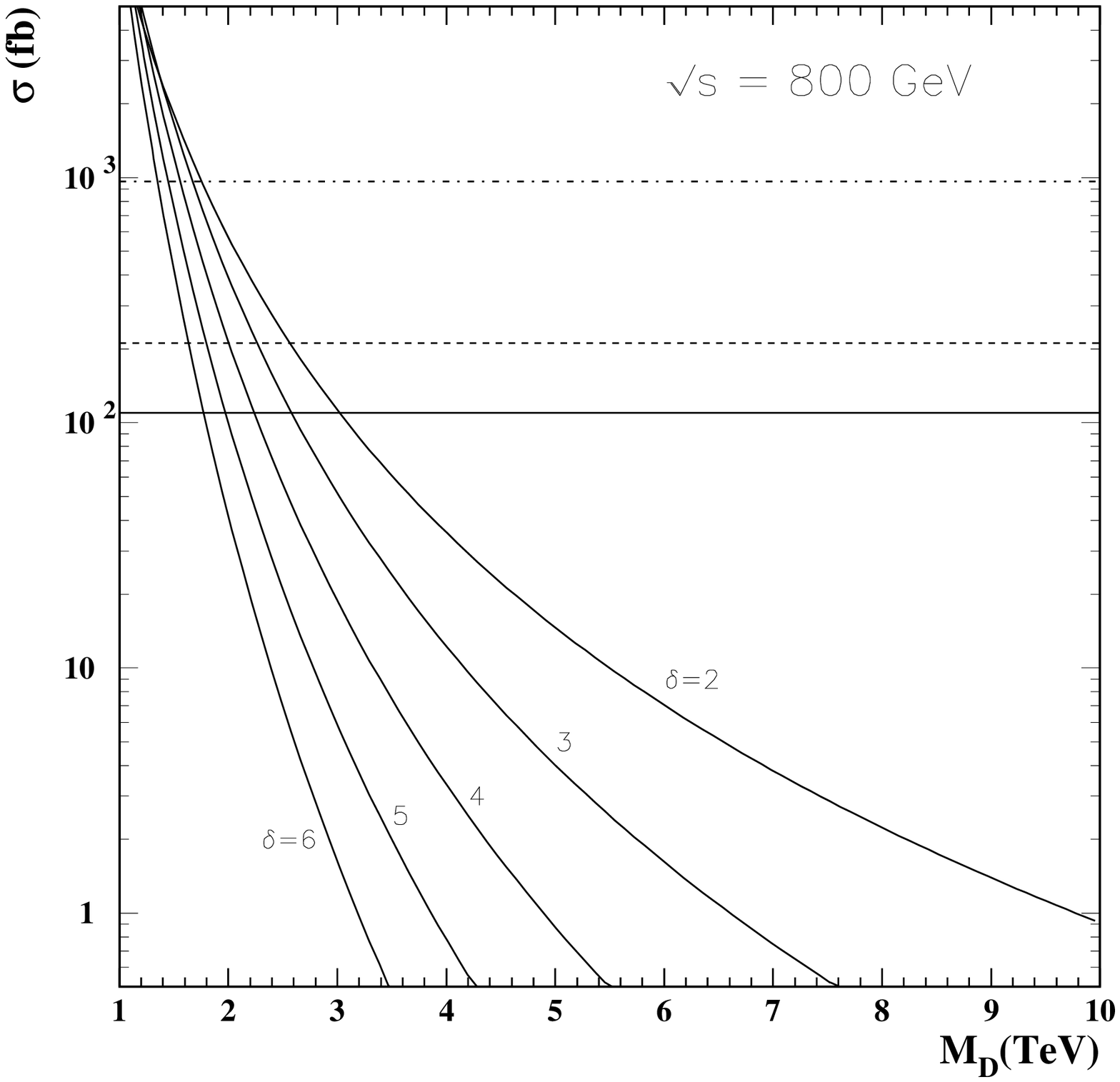}
\caption{Total cross section for $e^{+}e^{-} \rightarrow \gamma G^{(n)}$ 
at $\sqrt{s}=800$ GeV as a function of the scale $M$ (denoted as $M_{D}$ 
on the plot) for different number $d \equiv \delta$ of extra dimensions. The 
three horizontal lines indicate the background cross sections from 
$e^{+}e^{-} \rightarrow \gamma \nu \bar{\nu}$ for both beams 
polarized (solid line), only electron beam polarization (dashed) 
and no polarization (dot-dashed) \cite{TDR-Wil}.}
\label{fig:ADD-ee-M}
\end{figure}

At $e^{+}e^{-}$ colliders the main contribution comes from the 
$e^{+}e^{-} \rightarrow \gamma G^{(n)}$ process. 
The main background comes from the process 
$e^{+}e^{-} \rightarrow \nu \bar{\nu} \gamma$ and can be effectively 
suppressed by using polarized beams. Fig. \ref{fig:ADD-ee} shows 
the total cross section of the graviton production in 
electron-positron collisions \cite{ChKe}. Fig. \ref{fig:ADD-ee-M} shows the 
same cross section as a function of $M$ for $\sqrt{s} = 800 \; \mbox{GeV}$
\cite{TDR}, \cite{TDR-Wil}. 
 
In Table \ref{tab:ADD-ee-M} sensitivity in mass scale $M$ in 
TeV at 95\% C.L. is presented. The results for $\sqrt{s}=1$ TeV and 
the integrated luminocity  
${\cal L}= 200 \mbox{fb}^{-1}$ (left part of the table) are taken from 
Ref. \cite{GRW}. The expected sensitivity at TESLA (right part 
of the table) is taken from Ref.~\cite{TDR}. 
We see that in experiments with polarized beams the background is 
suppressed, and the upper value of $M$, for which the signal of the 
graviton creation can be detected, is higher.

\begin{table}[h]
\caption{}
\begin{center}
\begin{tabular}{||c||c|c|c|c|c||}
\hline
     & \multicolumn{2}{c|}{}  & \multicolumn{3}{c|}{TESLA:} \\  
 $d$ & \multicolumn{2}{c|}{$\sqrt{s}=1$ TeV, ${\cal L}= 200 \mbox{fb}^{-1}$} &
       \multicolumn{3}{c|}{$\sqrt{s}=800$ GeV, ${\cal L}= 1 \mbox{ab}^{-1}$} \\
\cline{2-6} 
 & $P_{-,+}=0$  & $P_{-,+}=0.9$  & 
 $P_{-,+}=0$ & $P_{-}=0.8$ & $P_{-,+}=0.8/0.6$ \\
 \hline 
$2$ & $4.1$ & $5.7$ & $5.9$ & $8.3$ & $10.4$ \\ \hline
$3$ & $3.1$ & $4.0$ & $4.4$ & $5.8$ & $6.9$ \\ \hline
$4$ & $2.5$ & $3.0$ & $3.5$ & $4.4$ & $5.1$ \\ \hline
\end{tabular}
\end{center} 
\label{tab:ADD-ee-M}
\end{table}

Effects due to gravitons can also be observed at hadron colliders. A  
characteristic process at the LHC would be $pp \rightarrow (\mbox{jet} + 
\mbox{missing} \; E)$. The subprocess which gives the largest 
contribution is the quark-gluon collision $qg \rightarrow qG^{(n)}$. Other 
subprocesses are $q\bar{q} \rightarrow gG^{(n)}$ and 
$gg \rightarrow gG^{(n)}$. The range of values of the scale 
$M$ (in TeV) which can lead to a discovery at at least $5\sigma$ 
for the direct graviton production at LHC (ATLAS study, Ref.~\cite{VaHi00}) 
and TESLA with polarized beams \cite{TDR} are presented in 
Table \ref{tab:ADD-rangeM} (the Table is borrowed from Ref.~\cite{TDR}). 
The lower value of $M$ appears because for $\sqrt{s} > M$ the 
effective theory approach breaks down. 
     
\begin{table}[h]
\caption{}
\begin{center}
\begin{tabular}{||c|c|c|c|c|c||}
\hline
$d$ & $2$ & $3$ & $4$ & $5$ & $6$ \\ 
 \hline
LHC & $4.0-7.5$ & $4.5-5.9$ & $5.0-5.3$ & none & none \\ 
\hline
TESLA & $0.5-7.9$ & $0.5-5.6$ & $0.5-4.2$ & $0.5-3.4$ & $0.5-2.9$ \\ 
\hline
\end{tabular}
\end{center}
\label{tab:ADD-rangeM}
\end{table}

Processes of another type, in which the effects of extra dimensions can be 
observed, are exchanges via virtual KK modes, namely 
virtual graviton exchanges. Contributions to the cross section from 
these additional channels lead to its deviation from 
the behaviour expected in the four-dimensional SM. The effect 
is similar to the one mentioned in Sect. 1.3 (see Figs.~\ref{fig:Delta-R}, 
\ref{fig:Delta-s}). Deviations due to the KK modes can be also observed   
in the left-right asymmetry. 
Since the momentum along the extra dimensions 
is not conserved on the branes, processes of such type appear already 
at tree-level. An example is  
$e^{+}e^{-} \rightarrow f \bar{f}$ with $G^{(n)}$ being the internal 
line. Moreover, gravitons can mediate processes absent in the SM 
at tree-level, for example, $e^{+}e^{-} \rightarrow HH$, 
$e^{+}e^{-} \rightarrow gg$. Detection of such events with the large 
enough cross section may serve as an indication of the existence 
of extra dimensions.  

The $s$-channel amplitude of a graviton-mediated scattering 
process is given by 
\[
{\cal A} = \frac{1}{M_{Pl}^{2}} \sum_{n} \left\{ 
T_{\mu \nu} \frac{P^{\mu \nu} P^{\rho \sigma}}{s-m_{n}^{2}} 
T_{\rho \sigma} 
+ \sqrt{\frac{3(d-1)}{d+2}} 
\frac{T^{\mu}_{\mu} T^{\nu}_{\nu}}{s-m_{n}^{2}}
\right\} = {\cal S} {\cal T},  
\]
where $P_{\mu \nu}$ is a polarization factor coming from the propagator 
of the massive graviton and $T_{\mu \nu}$ is the energy-momentum tensor 
\cite{GRW}. The factor ${\cal T}$ contains  
contractions of this tensor, whereas ${\cal S}$ is a kinematical factor, 
\bea
{\cal S} & = & \frac{1}{M_{Pl}^{2}} \sum_{n}\frac{1}{s-m_{n}^{2}} 
\approx \frac{1}{M_{Pl}^{2}} S_{d-1} \frac{M_{Pl}^{2}}{M^{d+2}} 
\int^{\Lambda} \frac{m^{d-1}dm}{s-m^{2}}  \nonumber \\
& = & \frac{S_{d-1}}{2M^{4}} 
\left\{ i\pi \left( \frac{s}{M^{2}} \right)^{d/2-1}  + 
\sum_{k=1}^{[(d-1)/2]} c_{k} \left( \frac{s}{M^{2}} \right)^{k-1} 
\left( \frac{\Lambda}{M} \right)^{d-2k} \right\}.   \nonumber
\eea
Here, as before, we substituted summation over KK modes by integration. 
Since the integrals are divergent for $d \geq 2$ the cutoff $\Lambda$ 
was introduced. It sets the limit of applicability of the 
effective theory. Because of the cutoff the amplitude cannot be calculated 
explicitly without knowledge of a full fundamental theory. 
Usually in the literature it is assumed that the amplitude is dominated 
by the lowest-dimensional local operator (see \cite{GRW}). This amounts 
to the estimate 
\[
{\cal A} \approx \frac{\lambda}{M^{4}} {\cal T} (\cos \theta), 
\]
where the constant $\lambda$ is supposed to be of order $1$.
Note that in this approximation ${\cal A}$ does not depend on the 
number of extra dimensions.  
Characteristic features are the spin-2 particle exchange 
and the gravitational universality. 

\begin{figure}[htbp]
\centerline{
\psfig{figure=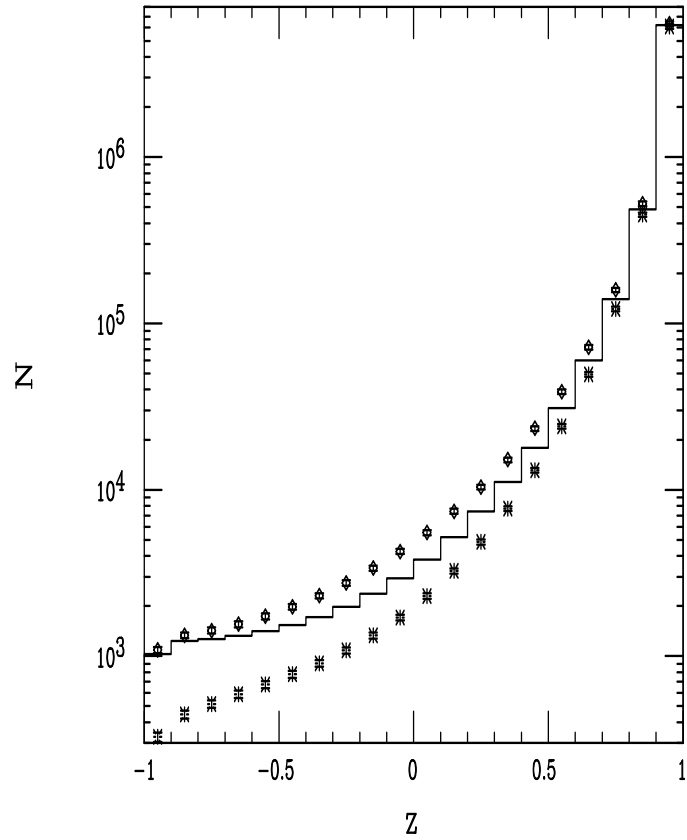,height=10.5cm,width=14cm,angle=-90}}
\vspace*{-0.9cm}
\caption{Deviation from the expectations of the SM (histogram) for
Bhabha scattering at a $500 \; \mbox{GeV}$ $e^{+}e^{-}$ collider
for the number $N$ of events per angular bin as a function of
$z=\cos \theta$ for $M = 1.5 \; \mbox{TeV}$ and the 
integrated luminocity ${\cal L}=75 \; \mbox{fb}^{-1}$ \cite{Ri99}. 
The two sets of data points correspond to the choices $\lambda = \pm 1$.} 
\label{fig:ADD-N}
\end{figure}

\begin{figure}[htbp]
\centerline{
\psfig{figure=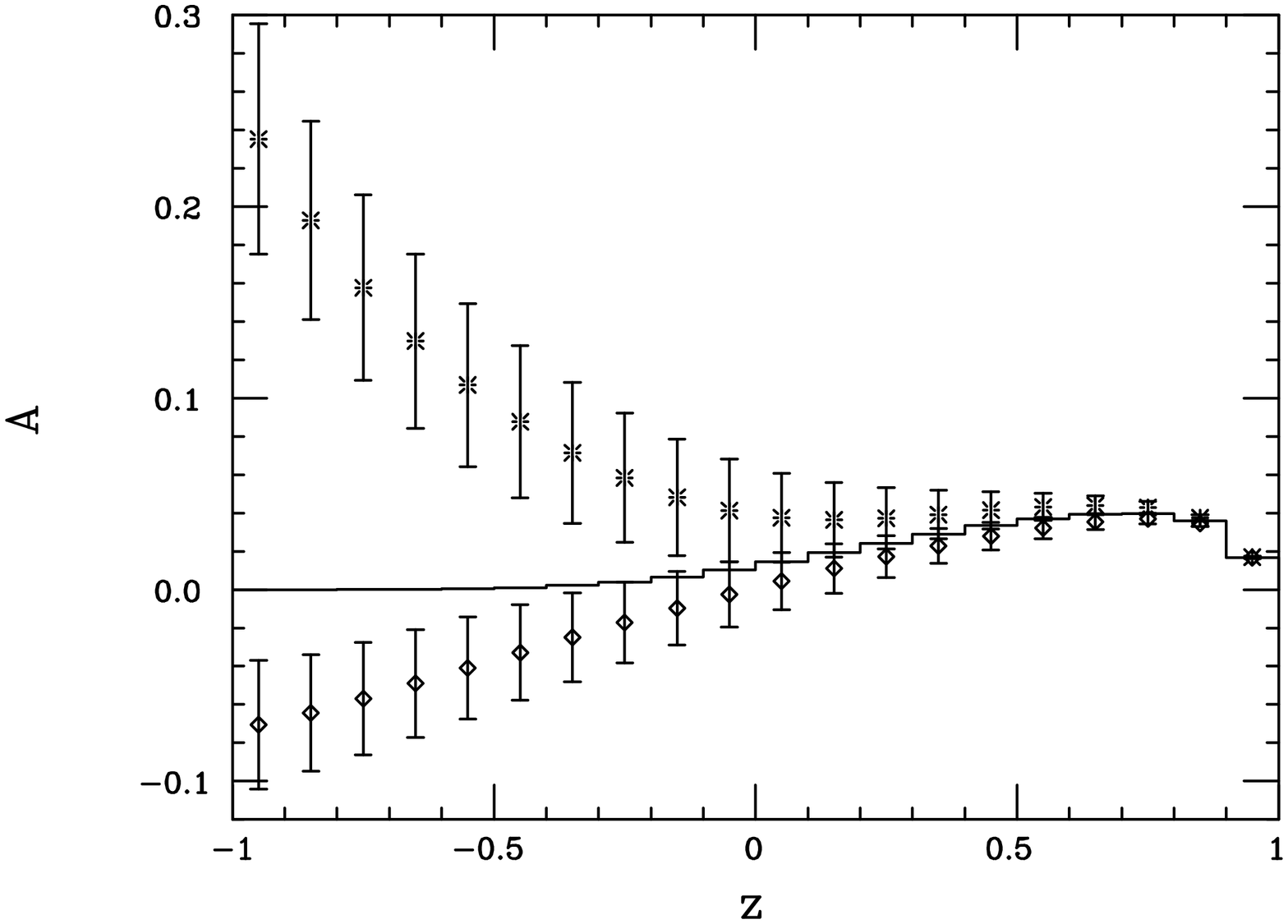,height=10.5cm,width=14cm,angle=-90}}
\vspace*{-0.9cm}
\caption{Deviation from the expectations of the SM (histogram) for
Bhabha scattering at a $500 \; \mbox{GeV}$ $e^{+}e^{-}$ collider
for the Left-Right polarization asymmetry as a function of
$z=\cos \theta$ for $M = 1.5 \; \mbox{TeV}$ and the 
integrated luminocity ${\cal L}=75 \; \mbox{fb}^{-1}$ \cite{Ri99}. 
The two sets of data points correspond to the choices $\lambda = \pm 1$.}
\label{fig:ADD-LRA}
\end{figure}

Typical processes, in which the virtual exchange via massive 
gravitons can be observed, are: (a) $e^{+}e^{-} \rightarrow \gamma \gamma$; 
(b) $e^{+}e^{-} \rightarrow f \bar{f}$, for example the Bhabha scattering 
$e^{+}e^{-} \rightarrow e^{+}e^{-}$ or M\"oller scattering 
$e^{-}e^{-} \rightarrow e^{-}e^{-}$; (c) graviton exchange contribution to 
the Drell-Yang production. A signal of the KK graviton mediated 
processes is the deviation in the number of events and in 
the left-right polarization asymmetry from 
those predicted by the SM (see Figs.~\ref{fig:ADD-N}, \ref{fig:ADD-LRA})  
\cite{Ri99}. 

The constraints on $M$ are essentially independent of $d$, 
the number of extra dimensions, and details of the fundamental theory. 
The spin-2 nature of the KK gravitons, mediating these processes,  
is revealed through angular distributions in $e^{+}e^{-}$ collisions in a 
unique way and can be distinguished from other sources of new 
physics. Signals for an exchange of the KK tower of gravitons appear in many 
complementary channels simultaneously.

As an illustration let us present combined estimates for the 
sensitivity in $M$ at 95\% C.L. obtained in Ref.~\cite{TDR} for TESLA 
by considering various processes:  

\begin{table}[h]
\begin{center}
\begin{tabular}{cc}
  $\sqrt{s} = 0.5$ TeV & $M \geq 5.6$ TeV \\
  & \\ 
  $\sqrt{s} = 0.8$ TeV & $M \geq 8$ TeV \\
\end{tabular}  
\end{center}
\end{table}

\vspace*{-0.5cm}
\nn
\vspace*{0.4mm}
\vspace*{-0.5cm}
\nn
\vspace*{0.4mm}

\subsection{Cosmology and Astrophysical constraints}

In the ADD model a conceivable concept of space-time, where the Universe 
is born and evolves, exists for temperatures $T \lsim M$ only. 
Recall that the fundamental scale $M$ is assumed to be 
$M \gsim 1 \; \mbox{TeV}$. For the predictions of the 
Big-Bang Nucleosynthesis (BBN) scenario not to be spoiled within the 
ADD model the expansion rate of the universe during the BBN cannot be 
modified by more than $10 \%$ approximately. Therefore, it should be 
required that before the onset of the BBN at the time $t=t_{BBN}$ 
 
(1) the size $R$ of extra dimensions is stabilized to its 
present value and satisfy 
\[
  R H |_{t=t_{BBN}} \sim \frac{1 mm}{10^{10} cm} \sim 10^{-13},   
\]

(2) the influence of extra dimensions on the 
expansion of the 3-brane is negligible. 

\noindent Here $H$ is the Hubble parameter. As a consequence,  
there exists some maximal temperature 
$T_{*}$ in the Universe, often referred to as the "normalcy temperature". 
For $T < T_{*}$ the bulk is virtually empty and $R$ is fixed. Usually 
the normalcy temperature is associated with the temperature 
of the reheating $T_{RH}$. 

Implementing the constraints, mentioned above, in various astrophysical 
effects leads to bounds on the fundamental scale $M$. 
They are summarized in Table \ref{tab:ADD-astro}. Let us make short 
comments concerning these bounds.

\begin{table}[h]
\caption{}
\begin{center}
\begin{tabular}{|c||c||c|}
\hline
Nature of the constraint & $d=2$ & $d=3$ \\
\hline \hline
1. Cooling of the Universe \cite{ADD2} & $M \gsim 1$ TeV & $M \gsim 1$ TeV \\
\hline
2. Overclosure of the Universe \cite{Hns01} & $M \gsim 10$ TeV 
                                                      & $M \gsim 1$ TeV \\
\hline
3. SN1987A cooling \cite{CuPe99,Hnh00} & $M \gsim 50$ TeV & $M \gsim 5$ TeV \\
\hline
4. CDG radiation \cite{ADD2,HaSm99,HnsRa01} & $M \gsim 80 \div 100$ TeV &
$M \gsim 7$ TeV \\
\hline
\end{tabular}
\end{center}
\label{tab:ADD-astro}
\end{table}

\noindent \underline{1. Cooling of the Universe.} The radiation on the 
3-brane cools in two ways: (1) due to the expansion of the Universe, 
the standard mechanism; and (2) due to the production of gravitons, which
is a new channel of cooling present in the ADD model. The rates of
decrease of the radiation energy density $\rho$ are given by
\bea
 {\cal R}_{1} & \equiv & \left. \frac{d \rho}{dt}\right|_{\mbox{expansion}}
 \sim -3H\rho \sim -3 \frac{T^{2}}{M_{Pl}} \rho,  \nonumber \\
 {\cal R}_{2} & \equiv & \left. \frac{d \rho}{dt}\right|_{\mbox{evaporation}}
 \sim -\frac{T^{d+7}}{M^{d+2}}. \nonumber
\eea
respectively \cite{ADD2}. 
The requirement ${\cal R}_{2} < {\cal R}_{1}$ gives the bound
on the normalcy temperature: 
\beq
 T_{*} < M \left( \frac{M}{M_{Pl}} \right)^{\frac{1}{d+1}} \sim
 10^{\frac{6d-9}{d+1}} \; \mbox{MeV}\; \left( \frac{M}{1 \; \mbox{TeV}}
 \right)^{\frac{d+2}{d+1}}.   \label{ADD:T-cool}
\eeq
For $d=2$ and $M = 1$TeV the bound is $T_{*} < 10 \; \mbox{MeV}$.
The lowest possible value of the temperature of reheating $T_{RH}$
acceptable within the BBN scenario is $T_{RH}=0.7 \; \mbox{MeV}$.
Hence, inequality (\ref{ADD:T-cool}) 
on $T_{*}$ does not set any stronger bound on $M$.

\begin{figure}[tp]
\epsfxsize=0.9\hsize
\epsfbox{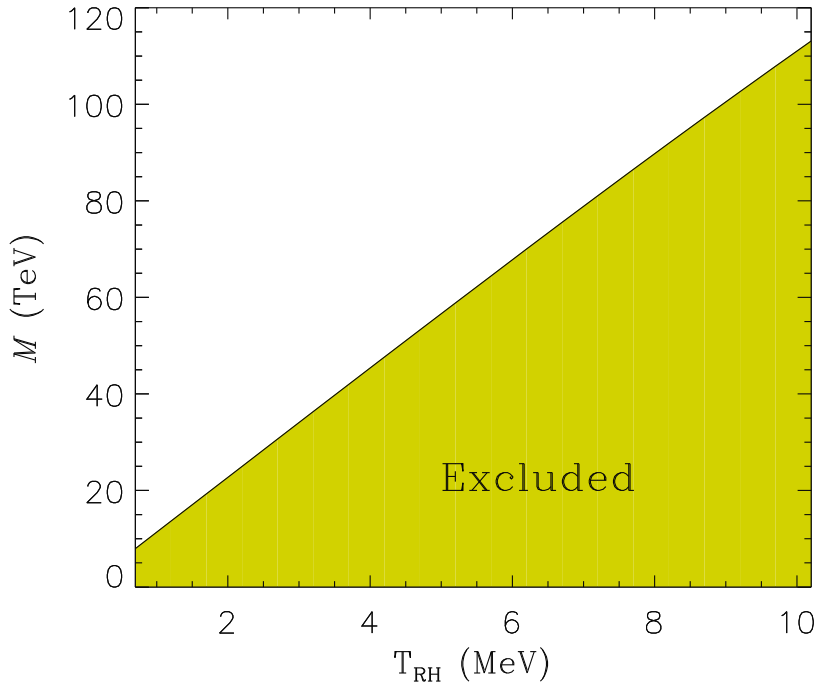}
\caption{The lower bound on $M$ as a function of the 
temperature of reheating $T_{RH}$, when only modes produced during the 
radiation dominated epoch are considered. The calculation is 
done for $d=2$ \cite{Hns01}.}
\label{fig:ADD-M-TRH}
\end{figure}

\noindent \underline{2. Overclosure of the Universe.}
Since the KK excitations are massive and their lifetime
$\tau_{grav} > \; (\mbox{age of the Universe})$ (see estimate 
(\ref{ADD-tau})) they may overclose
the Universe. The condition for this not to occur is
$\rho_{grav} < \rho_{crit}$, where $\rho_{crit}$ is the critical density. 
Writing this inequality for $\rho/T^{3}$ and taking into account that 
this ratio is invariant, one gets 
\cite{ADD2}
\[
\frac{\rho_{grav}}{T^{3}} \sim \frac{(T_{*} n_{grav})}{T_{*}^{3}} \sim
\frac{1}{T_{*}^{3}} \frac{T_{*}^{d+5} M_{Pl}}{M^{d+2}} <
\frac{\rho_{crit}}{T^{3}} \sim 3 \cdot 10^{-9} \; \mbox{GeV}.
\]
This amounts to the following bound on the normalcy temperature:  
\beq
T_{*} \lsim 10^{\frac{6d-15}{d+2}} \; \mbox{MeV} \cdot 
\frac{M}{1 \; \mbox{TeV}}    \label{ADD:T-closure}
\eeq   
which already sets a stronger bound on $M$. Thus, for $d=2$ and $M=1$TeV 
inequality (\ref{ADD:T-closure}) gives $T_{*} \lsim 0.2 \; \mbox{MeV}$. 
For $T_{*}$ to be higher than $0.7 \; \mbox{MeV}$ the scale $M$ must be 
$M \gsim 4 \; \mbox{TeV}$. More accurate estimates
give bounds on $M$ as a function of the temperature of reheating.   
They are shown in Fig. \ref{fig:ADD-M-TRH} and summarized in 
Table~\ref{tab:ADD-astro}.  

\noindent \underline{3. SN1987A cooling.}
The massive gravitons are copiously produced inside active stars, 
in particular inside the supernova SN1987A. The KK graviton emission 
competes with neutrinos carrying energy away from the stellar 
interior. This additional channel of energy loss accelerates the cooling 
of the supernova and can invalidate the current understanding of the 
late-time neutrino signal. To avoid this potential discrepancy with the 
observational data the supernova emissivity ${\cal E}$, calculated 
as the luminosity per unit mass of the star, should 
satisfy the Raffelt criterion:
\beq
    {\cal E} \leq 10^{19} 
    \frac{\mbox{ergs}}{\mbox{g} \cdot \mbox{s}}.  \label{ADD:Raff-crit}
\eeq
Assuming that the graviton emission is 
dominated by two-body collisions and that the main contribution 
comes from the nucleon-nucleon bremsstrahlung $N N \rightarrow NNG^{(n)}$  
it was shown in Ref. \cite{ADD2} that Eq. (\ref{ADD:Raff-crit})
implies  
\[
   M \gsim 10^{\frac{30 - 9d}{2(d+2)}} \; \mbox{TeV}.  
\]
For $d=2$ this gives $M \geq 30 \; \mbox{TeV}$. More accurate estimates 
were obtained in Refs. \cite{CuPe99}, \cite{Hnh00} and are
presented in Table~\ref{tab:ADD-astro}.  

\noindent \underline{4. CDG radiation.}
Though, according to Eq. (\ref{ADD-tau}), the lifetime of the 
massive KK mode exceeds the age of the Universe, the decay 
$G^{(n)} \rightarrow \gamma \gamma$ can still produce a noticeable 
effect in distorting the cosmic diffuse gamma (CDG) radiation. 

A simple estimate, obtained in Ref. \cite{ADD2}, gives 
\[
  T_{*} \lsim 10^{\frac{6d-15}{d+5}} \; \mbox{MeV} \cdot 
  \left( \frac{M}{1 \; \mbox{TeV}} \right)^{\frac{d+2}{d+5}}. 
\]
In the case of two extra dimensions the scale $M$ must satisfy  
$M \gsim 3 \; \mbox{TeV}$ in order to make possible  
the normalcy temperature to be $T_{*} \leq 0.7 \; \mbox{MeV}$.

A more detailed study of photons produced in the decay of KK gravitons 
was carried out in Ref.~\cite{HaSm99}. The lower limit on $M$ was 
obtained by comparing the result of calculation of the photon 
spectrum $dn_{\gamma}/dE \propto (M/\mbox{TeV})^{-(d+2)}$ with the 
upper bound on the recently measured value (COMPTEL results, 
see Ref.~\cite{COMPTEL}): 
\beq
\left. \frac{d n_{\gamma}}{dE} \right|_{T=T_{*}} \leq 
\left( \frac{dn_{\gamma}}{dE} \right)_{\mbox{exp}} \approx 
10^{-3} \; \mbox{MeV}^{-1} \cdot \mbox{cm}^{-2} \cdot \mbox{s}^{-1} 
\cdot \mbox{ster}^{-1}.    \label{ADD:CDG1}
\eeq
For $d=2$, the photon energy $E \simeq 4 \; \mbox{MeV}$, that gives 
the most stringent limit, and $T_{*}=1 \; \mbox{MeV}$ one gets 
$M > 110 \; \mbox{TeV}$.  

The strongest bounds on $M$ were obtained in Ref.~\cite{HnsRa01},  
they are presented in Table~\ref{tab:ADD-astro}. These bounds were derived 
by constraining the present-day contribution of the KK emission 
from supernovae, exploded in the whole history of the Universe, to 
the MeV $\gamma$-ray background. 

\vspace{0.3cm}

Scenarios of early cosmology within the ADD model can be very different 
from the standard ones. We will restrict ourselves to a few general remarks.  
First of all, for the effective theory 
to be valid the energy density of KK gravitons must satisfy 
\beq
    \rho_{KK} \ll M^{4}.    \label{ADD:rho-M}
\eeq
In addition to this, there is another constraint derived in Ref. 
\cite{KalLi98} which comes from the condition that the total 
energy in the bulk should be much less than the total energy on the wall. 
It ensures that the influence of the bulk on the expansion of 
the Universe can be neglected. It amounts to the inequality 
\[
\rho_{bulk} \ll M^{4+d} \left( \frac{M^{2}}{M_{Pl}^{2}} \right). 
\]

Another new feature is that the 
only natural time scale for the beginning of the inflation 
in the ADD model is $t_{0} \sim M^{-1}$. Using condition 
(\ref{ADD:rho-M}) the Hubble parameter at that time can be estimated 
as 
\[
H \sim \sqrt{\frac{8\pi V(\phi)}{3 M_{Pl}^{2}}} \lsim \frac{M^{2}}{M_{Pl}}
\ll M. 
\] 
Hence, the inflation occurs at a time scale $t_{i} = H^{-1} \gg M^{-1}$.
     
Further details of ADD cosmology depend on particular models. For example, 
as it was shown in Ref.~\cite{Lyth98} (see also 
\cite{MoPePi00}), within scenarios in which the inflaton is a brane field, 
i.e. a field localized on the brane and not propagating in the bulk, 
its mass is tiny comparing to the scale $M$. Namely,  
\beq
\frac{m_{inf}}{M} \lsim \frac{M}{M_{Pl}} \sim 10^{-15}. 
   \label{ADD:infl}
\eeq
We see that such scenarios introduce a new hierarchy, and perhaps the 
SUSY is needed to protect the inflaton mass from 
quantum corrections. They also have problems in describing the 
density perturbations. 
In Ref.~\cite{MoPePi00} it was argued that scenarios with the inflaton 
being a bulk field are free of many of these problems.   

\section{Randall-Sundrum Models}
\label{RS:model}

\subsection{RS1 model}

The Randall-Sundrum (RS) models are models based on solutions for 
the five-dimensional background metric obtained by L. Randall 
and R. Sundrum in Refs.~\cite{RS1}, \cite{RS2}.  

We begin with the model which was proposed in the first of these two papers,  
Ref.~\cite{RS1}, and which was termed as the RS1 model. 
It provides a novel and 
interesting solution to the hierarchy problem. 

The RS1 model is a model of Einstein gravity 
in a five-dimensional space-time with the extra dimension being 
compactified to the orbifold $S^{1}/Z_{2}$. There are two 3-branes 
located at the fixed points $y=0$ and $y=\pi R$ of the orbifold, 
where $R$ is the radius of the circle $S^{1}$.    
The brane at $y=0$ is usually 
referred to as brane 1, whereas the brane at $y=\pi R$ is called brane 2.   

We denote space-time coordinates by $\hat{x}^{M}$, where $M=0,1,2,3,4$, 
the four-dimensional coordinates by $x^{\mu} \equiv \hat{x}^{\mu}$, 
$\mu = 0,1,2,3$, and $\hat{x}^{4} \equiv y$. Let $\hat{G}_{MN}(x,y)$ 
be the metric tensor of the multidimensional gravity. Then  
$g^{(1)}(x) = \hat{G}_{\mu \nu}(x,0)$ and 
$g^{(2)}(x) = \hat{G}_{\mu \nu}(x,\pi R)$ describe the metrics induced 
on brane 1 and brane 2 respectively. The action of the model 
is given by 
\bea
S & = & \int d^{4}x \int_{-\pi R}^{\pi R} dy \sqrt{-\hat{G}} \left\{ 
2 M^{3} {\cal R}^{(5)} \left[\hat{G}_{MN}\right] + \Lambda  \right\} 
\nonumber \\
& + & \int_{B_{1}} d^{4}x \sqrt{-g^{(1)}} \left( L_{1} - \tau_{1} \right) + 
\int_{B_{2}} d^{4}x \sqrt{-g^{(2)}} \left( L_{2} - \tau_{2} \right), 
           \label{RS:L}
\eea
where ${\cal R}^{(5)}$ is the five-dimensional scalar curvature, $M$ is 
a mass scale (the five-dimensional "Planck mass") and $\Lambda$ is the 
cosmological constant. $L_{j}$ is a matter Lagrangian and $\tau_{j}$ 
is a constant vacuum energy on brane $j$ $(j=1,2)$. 

A background metric solution in such system satisfies  
the Einstein equation 
\[
\sqrt{-\hat{G}} \left[ {\cal R}^{(5)}_{MN} - 
   \frac{1}{2} \hat{G}_{MN} {\cal R}^{(5)} \right] = 
   \frac{1}{4M^{3}} \Lambda \sqrt{-\hat{G}} \hat{G}_{MN} 
   - \frac{1}{4M^{3}} T_{MN}.  
\]
If contributions of matter on the branes are neglected, then 
the energy-momentum tensor is determined by the vacuum energy terms: 
\beq
T_{MN} = \tau_{1} \sqrt{-g^{(1)}} g^{(1)}_{\mu \nu} 
         \delta_{M}^{\mu} \delta_{N}^{\nu} \delta (y) +  
         \tau_{2} \sqrt{-g^{(2)}} g^{(2)}_{\mu \nu} 
         \delta_{M}^{\mu} \delta_{N}^{\nu} \delta (y- \pi R). 
         \label{RS:T}
\eeq
The RS background solution describes the space-time with non-factorizable
geometry and is given by
\beq
ds^{2} = e^{-2\sigma (y)} \eta_{\mu \nu} dx^{\mu} dx^{\nu} + dy^{2}.
         \label{RS:RS1}
\eeq
Inside the interval $-\pi R < y \leq \pi R$ the function 
$\sigma (y)$ in the warp factor $\exp (-2\sigma)$ is equal to
\beq
   \sigma (y) = k |y|, \; \; \; (k > 0) \label{RS:sigma}
\eeq
and for the solution to exist the parameters must be fine-tuned to
satisfy the relations
\[
\tau_{1} = - \tau_{2} = 24 M^{3} k, \; \; \; \Lambda = 24 M^{3} k^{2}.
\]
Here $k$ is a dimensional parameter which was introduced for convenience. 
This fine-tuning is equivalent to the usual cosmological constant problem.
If $k > 0$, then  the tension on brane 1 is positive, 
whreas the tension $\tau_{2}$ on brane 2 is negative.

For a certain choice of the gauge the most general perturbed metric
is given by
\[
ds^{2} = e^{-2k|y|} \left( \eta_{\mu \nu} + \tilde{h}_{\mu \nu}(x,y) \right)
dx^{\mu} dx^{\nu} + (1 + \phi (x) ) dy^{2}.
\]

The Lagrangian of quadratic fluctuations
is not diagonal in $\tilde{h}_{\mu \nu}(x,y)$ and $\phi (x)$, 
it contains $\tilde{h}_{\mu \nu} \phi$
cross-terms. Correspondingly, the equations of motion, which are obtained
by expanding the Einstein equations around background solution  
(\ref{RS:RS1}) are coupled. The problem of diagonalization of the 
Lagrangian was considered in Ref.~\cite{PiRaZa00}, \cite{AIMVV}. 
Representing 
$\tilde{h}_{\mu \nu}(x,y)$ as a certain combination of a new field variable 
$h_{\mu \nu}$ and $\phi$ and using the freedom of the residual 
gauge transformations, the Lagrangian can be diagonalized and the 
equations can be decoupled. This procedure was carried out in a 
consistent way in Ref.~\cite{BKSV01}. 

As the next step the field $h_{\mu \nu}(x,y)$ is decomposed over 
an appropriate system of orthogonal and normalized functions: 
\beq
h_{\mu \nu} (x,y) = \sum_{n=0}^{\infty} h_{\mu \nu}^{(n)}(x)
  \frac{\chi_{n}(y)}{R},    \label{RS:decomp}
\eeq
where
\bea
\chi_{0}(y) & = & 2 \sqrt{kR} e^{-2k |y|},   \nonumber \\
\chi_{n}(y) & = & N_{n}\left[C_{1} Y_{2}
             \left(\frac{m_{n}}{k} e^{k|y|}\right) \right.
+ \left. C_{2} J_{2}\left(\frac{m_{n}}{k} e^{k|y|}\right)\right]
\; \; \; (n \neq 0). \nonumber
\eea
Here $J_{2}$ and $Y_{2}$ are the Bessel functions, $N_{n}$ are 
normalization factors. The boundary conditions 
(or junction conditions) on the branes,  
that are due to the $\delta$-function terms (see Eq. (\ref{RS:T})), fix the 
constants $C_{1}=Y_{1}(m_{n}/k)$ and $C_{2}=-J_{1}(m_{n}/k)$ and 
lead to the eigenvalue equation  
\[
 J_{1}\left( \beta_{n} e^{-k\pi R}\right) Y_{1}\left(\beta_{n}\right) - 
 Y_{1}\left(\beta_{n}e^{-k\pi R} \right) J_{1}\left( \beta_{n} \right) = 0. 
\]
The numbers $\beta_{n}$ are related to $m_{n}$ by 
$m_{n}=\beta_{n} k e^{-\pi k R}$. 
For small $n \geq 1$ this equation reduces to the approximate one: 
$J_{1} (\beta_{n}) = 0$, and $\beta_{n}$'s are equal to 
$\beta_{n} = 3.83, 7.02, 10.17, 13.32, \ldots$
for $n=1,2,3,4, \ldots$. The zero mode field $h_{\mu \nu}^{(0)}(x)$
describes the massless graviton. Within the
five-dimensional picture it appears as a state localized on brane 1.
The fields $h_{\mu \nu}^{(n)}(x)$ with $n \geq 1$ describe massive KK modes.
The field $\phi (x)$ classically satisfies the equation of
motion for a scalar massless field: $\Box_{(4)} \phi (x) = 0$. This field
was called radion. It represents the degree of freedom corresponding
to the relative motion of the branes. Apparently, the radion as a
particle was first identified in Ref.~\cite{AHDMR01} (see also 
\cite{Sund98}) and studied in articles~\cite{Cs99} - \cite{ChGrRu99}.

As we will see later, it is brane 2 which is most interesting from the
point of view of the high energy physics phenomenology. However,
because of the non-trivial warp factor $e^{-2\sigma (\pi R)}$ on brane 2,
the coordinates $\{ x^{\mu}\}$ are not Galiliean (coordinates are
called Galilean if $g_{MN}= diag(-1,1, \dots,1)$). To have the correct
interpretation of the effective theory on brane 2 one introduces
the Galilean coordinates $z^{\mu}=x^{\mu} e^{-\pi k R}$. Correspondingly, 
the gravitational field and the energy-momentum tensor should be rewritten in
these coordinates. Calculating the zero mode sector of the effective
theory we obtain
\[
S_{eff} \supset \int d^{4}z e^{4k\pi R}
\int_{-\pi R}^{\pi R} dy e^{-4k|y|} \left\{ 2M^{3} e^{-2k(\pi R-|y|)}
{\cal R}^{(4)} \left[\eta_{\mu \nu}+ h_{\mu \nu}^{(0)}\right] + 
\ldots \right\}.
\]
Identifying the coefficient, which multiplies the four-dimensional
scalar curvature, as $M_{Pl}^{2}$ one gets
\beq
M_{Pl}^{2} = e^{2k\pi R} \int_{-\pi R}^{\pi R} dy e^{-2k|y|} =
\frac{M^{3}}{k} \left( e^{2k \pi R} - 1 \right)    \label{RS:M-Pl}
\eeq
\cite{Rub01}, \cite{BKSV01}, \cite{GNS01}.
In the Galilean coordinates the four-dimensional 
effective action of quadratic
fluctuations after expansion over KK modes is given by 
\[
S_{eff} = \int d^{4}z \left\{ -\frac{1}{4} \sum_{n=0}^{\infty}
\left[ \partial_{\mu} h^{(n)}_{\rho \sigma} \partial^{\mu} h^{(n)\rho \sigma}
+ (m_{n} e^{k\pi R} )^{2} h^{(n)}_{\rho \sigma} h^{(n)\rho \sigma} \right]
- \frac{1}{2} \partial_{\mu} \varphi \partial^{\mu} \varphi \right\}.
\]
Here the indices are raised with the Minkowski metric $\eta_{\mu \nu}$,
and the field $\varphi (x)$ is related to $\phi$ by the rescaling
\[
\phi = \sqrt{\frac{e^{2kR}-1}{3kR^{2}} } \varphi 
\]
which is chosen so that to bring the kinetic term of the radion to 
the canonical form.  
The masses of KK modes are equal to $M_{n} = m_{n}e^{k \pi R} = k \beta_{n}$.

The general form of the interaction
of the fields, emerging from the five-dimensional metric,  with the matter  
localized on the branes is given by the expression:
\[
\frac{1}{2 M^{3/2}} \int_{B_{1}} d^{4}x h_{\mu \nu}(x,0) T^{(1)}_{\mu \nu} +
\frac{1}{2 M^{3/2}} \int_{B_{2}} d^{4}z h_{\mu \nu}(z,0) T^{(2)}_{\mu \nu}
\sqrt{-\det \gamma_{\mu \nu} (\pi R)} 
\]
following from Eq. (\ref{RS:L}). Here $T^{(1)}_{\mu \nu}$ 
and $T^{(2)}_{\mu \nu}$ are the energy-momentum
tensors of the matter on brane 1 and brane 2 respectively. Decomposing
the field $h_{\mu \nu}(x,y)$ according to (\ref{RS:decomp}) 
and rescaling the radion field, we obtain that
the interaction term on brane 2 is equal to
\beq
\frac{1}{2} \int_{B_{2}} d^{4}z \left[\kappa h^{(0)}_{\mu \nu} T^{(2)\mu \nu}
- \sum_{n=1}^{\infty} \kappa_{n} h^{(n)}_{\mu \nu} T^{(2)\mu \nu}
- \frac{\kappa_{rad}}{\sqrt{3}}T^{(2)\mu}_{\mu} \right],  \label{RS:int1} 
\eeq
where
\bea
\kappa & = & \frac{\sqrt{k}}{M^{3/2}}
\frac{e^{-k \pi R}}{\sqrt{1-\exp (-2k \pi R)}}, \label{RS:const} \\
\kappa_{n} & = & \kappa w_{n} e^{k \pi R}, \; \; \;
\kappa_{rad} = \kappa e^{k \pi R}.               \nonumber
\eea
The factors $w_{n}$ are determined by the values $\chi_{n}(\pi R)$
of the eigenfunctions at $y=\pi R$. One can check that
$w_{n} \approx 1.0$ for small $n$.

If a few first massive KK gravitons have masses $M_{n} \sim 1$TeV, then
a new interesting phenomenology in TeV-region of energies takes 
place on brane 2. To have this situation the fundamental mass scale 
$M$ and the parameter $k$ are taken to be $M \sim k \sim 1$TeV. 
Then, to satisfy relation
(\ref{RS:M-Pl}) we choose the radius $R$ such that $k R \approx 12$.
With this choice $e^{k\pi R} \sim 10^{15}$, thus giving a solution
to the hierarchy problem, namely allowing to relate two different scales: the
TeV-scale and the Planck scale. In this model the 
Planck scale is generated from
the TeV-scale via the exponential factor and no new large hierarchies are
created. The exponenial factor is of geometrical origin: it comes from the
warp factor of the RS solution.

The zero mode $h^{(0)}_{\mu \nu}(x)$ describes the usual massless graviton. 
Its coupling $\kappa$ to matter in Eq. (\ref{RS:int1}) is therefore 
identified with $M_{Pl}^{-1}$. This is consistent
because, as one can easily see, Eq. (\ref{RS:const})  
with $\kappa = M_{Pl}^{-1}$ gives the same relation (\ref{RS:M-Pl}).
The interaction Lagrangian can be rewritten in the following way:
\beq
\frac{1}{2} \int_{B_{2}} d^{4}z \left[
\frac{1}{M_{Pl}} h^{(0)}_{\mu \nu}(z) T^{(2)\mu \nu}
- \sum_{n=1}^{\infty} \frac{w_{n}}{\Lambda_{\pi}} h^{(n)}_{\mu \nu}
T^{(2)\mu \nu}
- \frac{1}{\Lambda_{\pi}\sqrt{3}}T^{(2)\mu}_{\mu} \right],   \label{RS:int2}
\eeq
where $\Lambda_{\pi} = M_{Pl} e^{-k \pi R} \approx \sqrt{M^{3}/k}$. We see
that the effective theory on brane 2 describes the massless graviton
(spin-2 field), the massless radion (scalar field) and the infinite
tower of massive spin-2 fields (massive gravitons) with masses
$M_{n} = k\beta_{n}$. The massless graviton, as in the standard gravity,
interacts with matter with the coupling $M_{Pl}^{-1}$. The interaction
of the massive gravitons and radion are considerably stronger: their
couplings are $\propto \Lambda_{\pi}^{-1} \sim 1 \; \mbox{TeV}^{-1}$.
This leads to new effects which in principle can be seen at the future
colliders. In the literature brane 2 (the brane with negative tension) is
referred to as the TeV-brane.

We would like to note that instead of choosing the Ga\-li\-li\-ean
coordinates on the TeV-brane, one can
introduce global five-dimensional coordinates $\{ z^{\mu},y \}$ with
$z^{\mu}=x^{\mu} e^{-k(\pi R - y)}$, for which
the warp factor on this brane is equal to 1.
This alternative description was used in Ref. \cite{Rub01}. All physical
conclusions, derived there, are the same as the ones obtained within
the formalism with the physical (Galilean)
coordinates on the TeV-brane described above, see
Refs. \cite{BKSV01}, \cite{GNS01}.
In the latter formalism the effective Lagrangian and formulas 
(\ref{RS:const}), expressing the coupling constants in terms of 
the parameters $M$, $k$ and $R$ of the model, differ from the
ones usually used in the literature (see, for example, Refs.
\cite{RS1,DHR1,DHR2}). In particular, we get $ k \sim 1$TeV, whereas in the
above mentioned  papers one needs $k \sim M_{Pl}$\footnote{The
alternative of choosing $M \sim k \sim 1 \; \mbox{TeV}$ was mentioned
in Ref. \cite{RS1}}. Nevertheless, it is easy to check that the
ratios $\kappa_{n}/\kappa$ and $\kappa_{rad}/\kappa$
are the same both in  physical (Galilean) coordinates $\{ z^{\mu} \}$
and in "non-physical" coordinates $\{ x^{\mu} \}$. For this reason
phenomenological predictions obtained in these and many other
previous papers remain valid.

To conclude this subsection let us discuss briefly two other RS-type models.
In article \cite{RS2} a model with Lagrangian (\ref{RS:L}) and non-compact
fifth dimension was proposed. This model contains only one brane, at $y=0$, 
and the fields of the SM are supposed to be localized there. 
The extra dimension is the infinite line. This model was called the RS2 model. 

The solution for the background metric
is given by the same Eqs. (\ref{RS:RS1}), (\ref{RS:sigma}), but now they are 
valid for $-\infty < y < \infty$. Fluctuations around the 
solution include a state with zero mass, which describes the massless 
graviton, and massive states. The massless graviton 
is localized on the brane, hence no contradiction with the Newton law appears 
at distances $r \gg k^{-1}$ with the parameter $k$ chosen to be 
$k \sim M_{Pl}$. Non-zero KK states are non-localized and form the continuous 
spectrum starting from $m=0$ (no mass gap). The RS2 model gives an elegant 
example of localized gravity with non-compact extra dimension. However, 
it does not provide a solution of the hierarchy problem. 

The Lykken-Randall (LR) model, proposed in Ref.~\cite{LR99}, is a combination 
of the RS1 and RS2 models. This is a model (\ref{RS:L}) with the non-compact 
fifth dimension and with two branes. Brane 1, the Planck brane, is 
located at $y=0$ and its tension determines the same background solution 
for the metric as in the RS2 model. Brane 2, the TeV-brane, is regarded as a 
probe brane, i.e. the tension $\tau_{2} \ll \tau_{1}$, so that 
it does not affect the solution. The TeV-brane is located at $y=\pi R$, 
and the value of $R$ is adjusted in such a way that  
\[
  M_{Pl} e^{-k \pi R} \sim M_{Pl} \cdot 10^{-15} \sim 1 \; \mbox{TeV}. 
\]
This assures that the hierarchy problem is solved on the TeV-brane. 
Therefore, it is considered to be "our" brane, i.e. the brane where the 
SM is localized. Note that the tension $\tau_{2}$ can be chosen to be 
positive.  

\subsection{HEP phenomenology}

In this subsection we discuss some effects in the RS1 model which can 
be observed in collider experiments at TeV-energies. 
They were studied in Refs.~\cite{DHR1} - \cite{DHR2} and other 
subsequent papers. 

Let us recall that according to Eq. (\ref{RS:int2}) 
the couplings of the fields are 
determined by the Planck mass and $\Lambda_{\pi}$, which are related 
to the parameters of the model by 
\[
 M_{Pl}^{2} \approx \frac{M^{3}}{k} e^{2k\pi R}, \; \; \;
\Lambda_{\pi} = M_{Pl} e^{-k \pi R} \sim 1 \; \mbox{TeV}.   
\]
The presence of the massless scalar radion leads to a contradiction 
with the known experimental data. That is why it is assumed that 
the radion is stabilized by that or another mechanism and, thus, 
becomes massive. One of such mechanisms was proposed in 
Refs.~\cite{AHDMR01}, \cite{GolWi99}. 
It provides the mass $\sim (10 \div 100)\; \mbox{GeV}$ to the radion 
without strong fine-tuning of parameters. With such mass the radion  
does not violate the Newton gravity law on the TeV-brane. There is much 
literature on the phenomenology of the radion (see, for example, 
Refs.~\cite{GoWi99a}, \cite{HEP-radion} and references therein), 
we do not discuss it here. 

Processes at high energies in the RS1 model (excluding the radion sector) 
are completely determined by two parameters. It is common to 
choose them to be  
\[
M_{1}=\beta_{1} k = \beta_{1} \Lambda_{\pi} \cdot 
\left( \frac{k}{M_{Pl}} e^{k \pi R}\right), 
\]
the mass of the first mode, and $\eta = (k/M_{Pl})e^{k \pi R}$
\cite{DHR1}. Recall that $\beta_{1} \approx 3.83$. 
Indeed, it is easy to check that all the couplings and observables can 
be expressed in terms of $M_{1}$ and $\eta$. For example, the 
total width of the first graviton resonance is equal to 
$\Gamma_{G^{(1)}} = \rho M_{1} \beta_{1}^{2} \eta^{2}$, where 
$\rho$ is a constant which depends on the number of open decay channels 
\cite{DHR1}.  

According to the assumptions made in the previous subsection, $\eta \sim 1$.   
There are two more restrictions on this parameter.

(1) The five-dimensional scalar curvature calculated for the RS solution 
is equal to ${\cal R}^{(5)} = - 20 k^{2} e^{2k \pi R}$. 
The RS solution can be trusted if 
\beq
|{\cal R}^{(5)}| < M^{2}e^{2k\pi R}.  \label{RS:R5-M2}
\eeq
This gives the constraint 
$k/M < 0.22$ or 
\beq
\eta = \left( \frac{k}{M} \right)^{3/2} 
\frac{1}{\sqrt{1-e^{-2k\pi R}}} \approx 
\left( \frac{k}{M} \right)^{3/2} < 0.1.    \label{RS:k-M}
\eeq

(2) Within the four-dimensional heterotic string model it can be shown 
(see, for example, \cite{DHR1}) that 
\[
\eta \geq \frac{1}{(2 \pi)^{3/2}} 
\frac{g^{2}_{\mbox{\small YM}}}{\sqrt{24 g_{\mbox{\small string}}}}. 
\]
For the gauge constant $g_{\mbox{\small YM}} \sim 0.7$ and 
string constant $g_{\mbox{\small string}} \sim 1$ this relation 
gives the inequality $\eta \gsim 0.01$. Combining this bound with 
inequality (\ref{RS:k-M}) we arrive at the following conservative estimate:
\beq
0.01 \lsim \eta \lsim 0.1.     \label{RS:k-M1}
\eeq

For such values of the parameter $\eta$ and the mass of the 
first KK mode $M_{1} \sim 1 \; \mbox{TeV}$ direct searches of the first 
KK graviton $G^{(1)}$ in the resonance production at the future
colliders become quite possible. 
Signals of the graviton detection can be 

(a) an excess in Drell-Yang processes 
\bea
& & q \bar{q} \rightarrow G^{(1)} \rightarrow l^{+}l^{-},    
  \nonumber \\
& & gg \rightarrow G^{(1)} \rightarrow l^{+}l^{-};  \nonumber 
\eea
      
(b) an excess in the dijet channel
\[
q \bar{q}, gg \rightarrow G^{(1)} \rightarrow q \bar{q}, gg. 
\]

\begin{figure}[htbp]
\centerline{
\psfig{figure=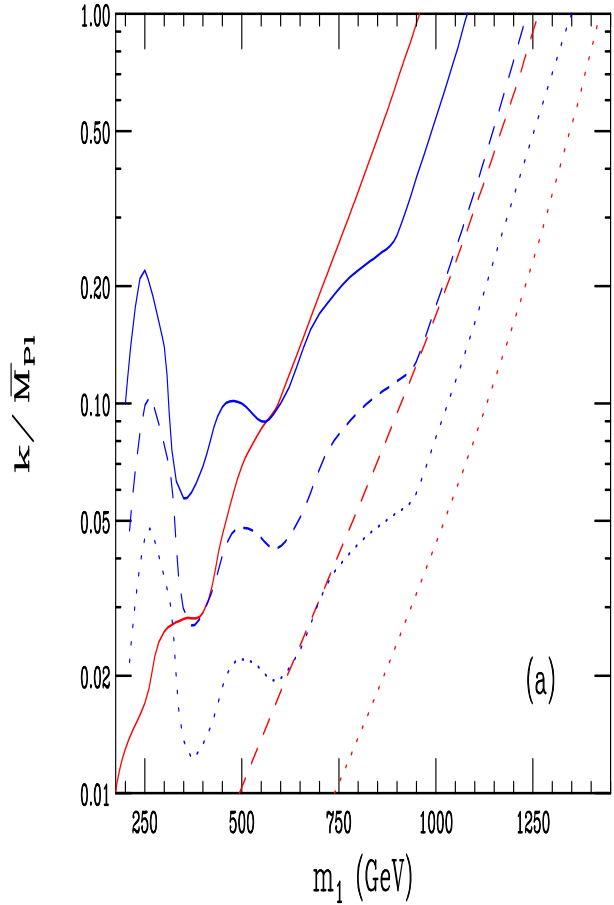,height=8.cm,width=12cm,angle=90}}
\vspace*{0.25cm}
\caption{Exclusion region for resonance production of the first KK graviton
excitation in the Drell-Yan (corresponding to the diagonal lines)
and dijet (represented by the bumpy curves) channels at the Tevatron.  
The solid curves represent the results
for Run I, while the dashed, dotted curves correspond to Run II with 2, 30
\infb\ of integrated luminosity, respectively.
The excluded region lies above and to the left of the curves.}
\label{fig:RS-Tev}
\end{figure}

\begin{figure}[htbp]
\centerline{
\psfig{figure=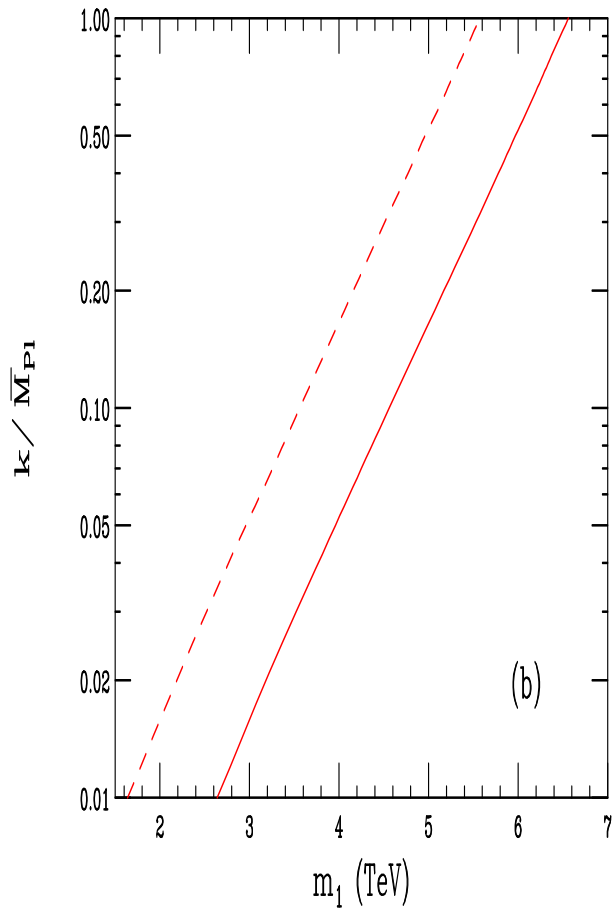,height=8.cm,width=12cm,angle=90}}
\vspace*{0.25cm}
\caption{Exclusion region for resonance production of the first KK graviton
excitation in the Drell-Yan production at the LHC. 
The dashed, solid curves correspond to
10, 100 \infb\ of integrated luminosity, respectively. 
The excluded region lies above and to the left of the curves.}
\label{fig:RS-LHC}
\end{figure}

\begin{figure}[htbp]
\centerline{
\psfig{figure=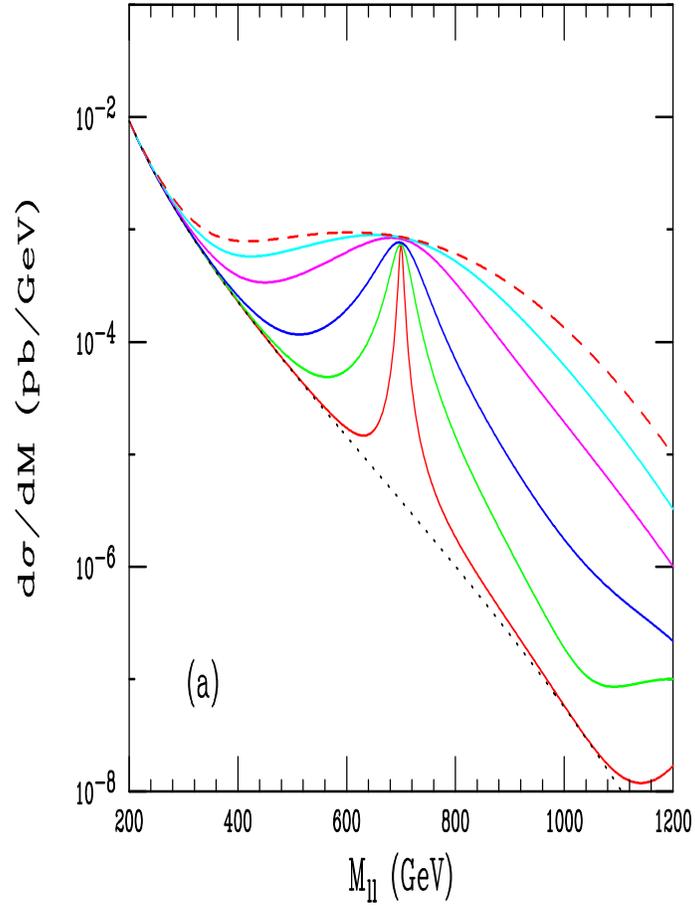,height=9.cm,width=12cm,angle=90}}
\vspace{0.1cm}
\caption{Drell-Yang production of the KK graviton with $M_{1}=700$ GeV
at the Tevatron for $\eta = 1,0.7,0.5,0.3,0.2$, and $0.1$, respectively,
from top to bottom \cite{DHR1}.}
\label{fig:RS:DYcs1}
\end{figure}

\begin{figure}[htbp]
\centerline{
\psfig{figure=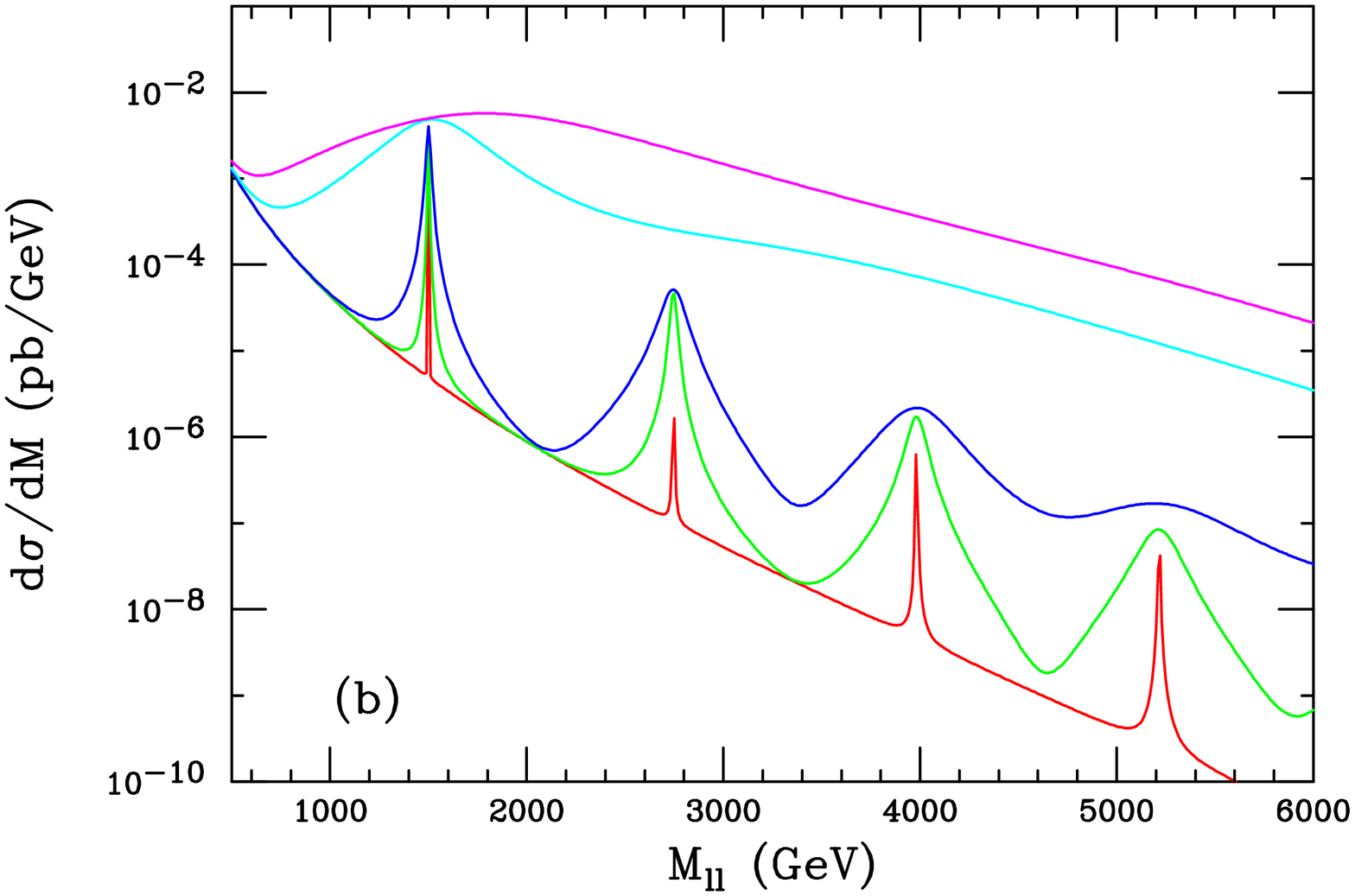,height=9.cm,width=12cm,angle=90}}
\vspace{0.1cm}
\caption{Drell-Yang production of the KK graviton with $M_{1}=1500$ GeV 
and its subsequent tower states at the
LHC for $\eta = 1,0.5,0.1,0.05,0.2$, and $0.01$, respectively, from top
to bottom \cite{DHR1}.}
\label{fig:RS:DYcs2}
\end{figure}

The plots of the exclusion regions for the Tevatron and LHC, taken from 
Ref.~\cite{DHR1}, are presented in Figs.~\ref{fig:RS-Tev}, 
\ref{fig:RS-LHC}.
The behaviour of the cross-section of the
Drell-Yang process as a function of the invariant mass of the final 
leptons for two values of $M_{1}$ and a few values of $\eta$ are shown in 
Figs.~\ref{fig:RS:DYcs1}, \ref{fig:RS:DYcs2}.    
\nn
\nn

\begin{figure}[tp]
\epsfxsize=0.9\hsize
\epsfbox{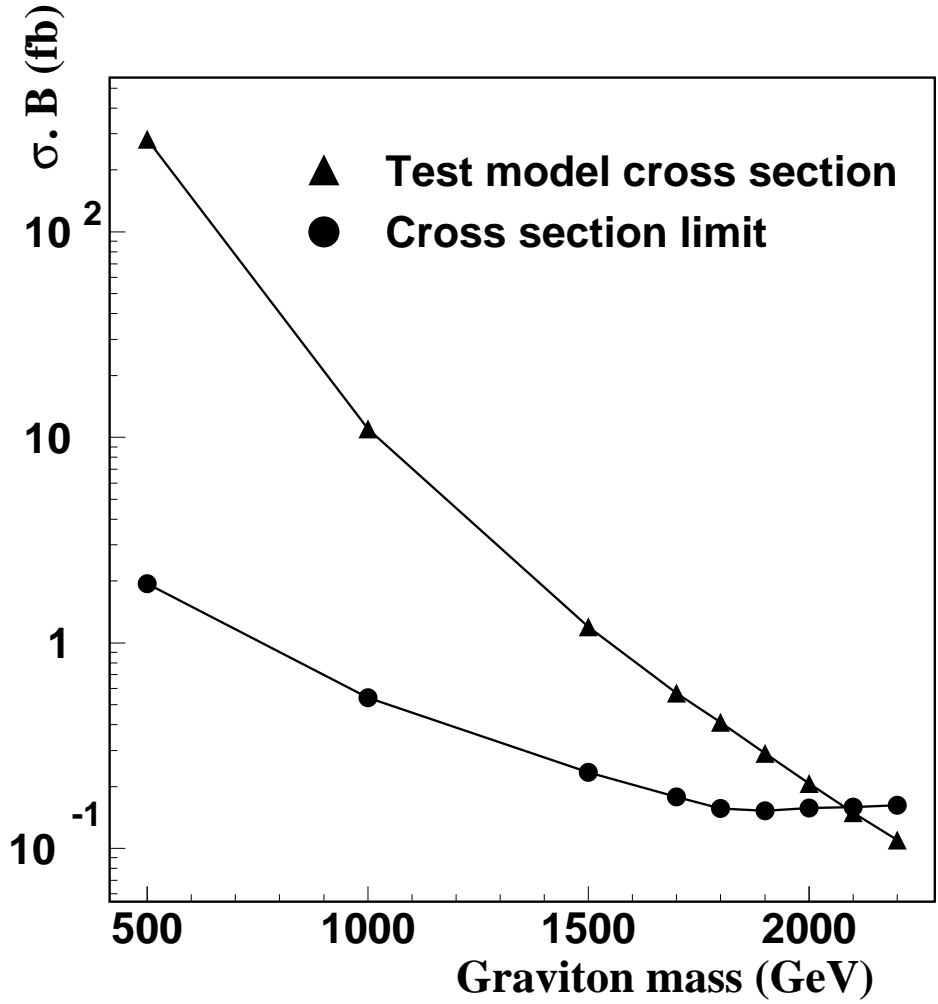}
\caption{The cross-section times branching ratio, $\sigma \cdot B$, 
for $G^{(1)} \rightarrow e^{+}e^{-}$ in the RS1 model and the smallest 
detectable cross-section times the branching ratio, $(\sigma \cdot B)^{min}$ 
\cite{AOPW00}. The plot is calculated for $\eta = 0.01$.}
\label{fig:RS:G1cs}
\end{figure}

As an example let us discuss the possibility of detection of the 
resonance production of the first massive graviton in the proton - 
proton collisions $p p \rightarrow G^{(1)} \rightarrow e^{+}e^{-}$ 
at the LHC (ATLAS experiment) 
studied in Ref.~\cite{AOPW00}. The main background processes are  
$p p \rightarrow Z/\gamma^{*} \rightarrow e^{+}e^{-}$.
By estimating the cross section of $G^{(1)} \rightarrow 
e^{+}e^{-}$ as a function of $M_{1}$ it was shown 
that the RS1 model would be detected if $M_{1} \leq 2080 \; \mbox{GeV}$, 
see Fig.~\ref{fig:RS:G1cs}.
 
To be able to conclude that the observed resonance is a graviton and not, 
for example, a spin-1 $Z'$ or a similar particle it is necessary to check 
that it is produced by a spin-2 intermediate state. The spin of the 
intermediate state can be determined from the analysis of the 
angular distribution function $f(\theta)$ of the process,
where $\theta$ is the angle between the initial and final beams.  
This function is $f=1$ for the scalar resonance, 
$f(\theta) = 1 + \cos^{2} \theta$ for a vector resonance, and 
is a polynomial of the 4th order in $\cos \theta$ for a spin-2 resonance. 
For example, for $gg \rightarrow G^{(1)} \rightarrow e^{+}e^{-}$  
$f(\theta) = 1 - \cos^{4} \theta$, whereas for $q\bar{q} \rightarrow G^{(1)} 
\rightarrow e^{+}e^{-}$  $f(\theta) =  1 - 3 \cos^{2} \theta + 
4\cos^{4} \theta$. The analysis, carried out in Ref.~\cite{AOPW00}, 
shows that angular distributions allow to determine the spin of the 
intermediate state with 90\% C.L. for $M_{1} \leq 1720$ GeV. 

As a next step it would be important to check the universality of the 
coupling of the first massive graviton $G^{(1)}$ by studing 
various processes, e.g. 
$pp \rightarrow G^{(1)} \rightarrow l^{+}l^{-}, \; \mbox{jets}, \; 
\gamma \gamma, W^{+}W^{-}, HH$, etc. If it is kinematically feasible to 
produce higher KK modes, measuring the spacings of the spectrum 
will be another strong indication in favour of the RS1 model. 

\begin{figure}[htbp]
\centerline{
\psfig{figure=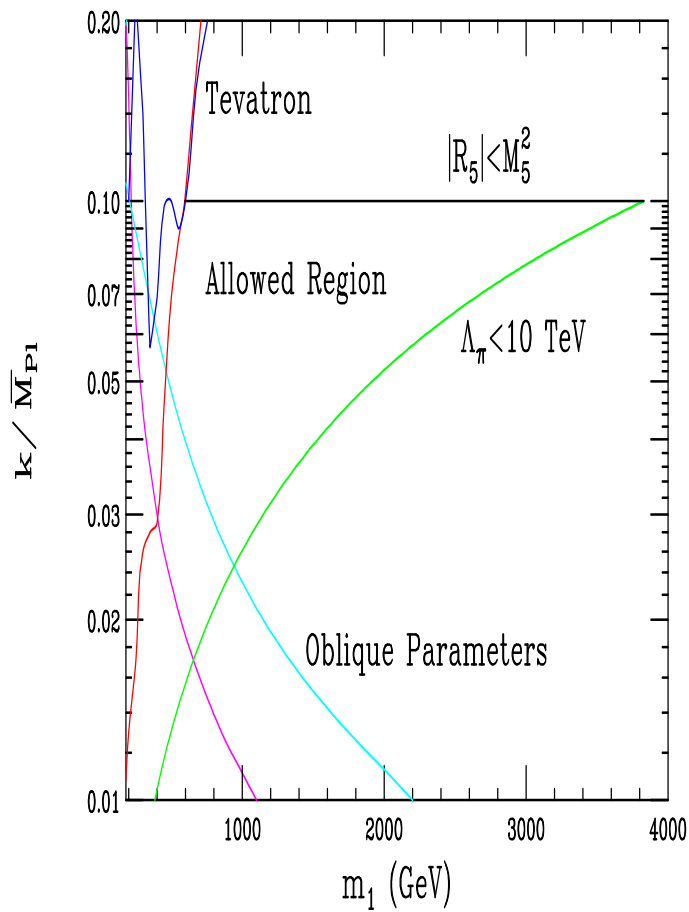,height=9cm,width=12cm,angle=90}}
\caption{Summary of experimental and theoretical 
constraints on the parameters $M_{1}$ and $\eta = (k/M_{Pl}) e^{k\pi R}$ 
of the RS1 model ($M_{1}$ is denoted as $m_{1}$ and $k/\bar{M}_{Pl}$ 
stands for $\eta$) \cite{DHR2}. The allowed region lies in the
center as indicated.}
\label{fig:RS:a.r.}
\end{figure}

The conclusion drawn in Ref.~\cite{DHR2} is that with the integrated 
luminocity 
${\cal L} = 100 \; \mbox{fb}^{-1}$ the LHC will be able to 
cover the natural region of parameters $(M_{1},\eta)$ and, therefore, 
discover or exclude the RS1 model. This is 
illustrated in Fig.~\ref{fig:RS:a.r.}. The curves represent 
the theoretical constraint on the scalar curvature, Eq.~(\ref{RS:R5-M2}) 
($M_{5} = M e^{2k\pi R}$),  
the Tevatron bound (see Figs.~\ref{fig:RS-Tev}, \ref{fig:RS-LHC}), 
the global fit from measurements of the oblique parameters
S and T, and the bound $\Lambda_{\pi} < 10 \; \mbox{TeV}$. The latter 
is regarded as a condition for solving the hierarchy problem, i.e.   
it is supposed that if $\Lambda_{\pi}$ is large enough, 
namely $\Lambda_{\pi} \geq 10$TeV, the hierarchy remains and 
the motivation for introducing the RS1 model with 
one extra dimension is not sufficient. The range of the 
region in the $\eta$-direction is given by interval (\ref{RS:k-M1}).     

There is an interesting phenomenology in theories when gauge fields and/or 
fermions of the SM are allowed to propagate in the bulk. We do not 
consider such cases here and present only a list of main references 
on the subject, see Refs.~\cite{GhPo1}, \cite{DHR3}, \cite{DHR2}, 
\cite{RS-SMbulk} - \cite{HuSch00}.  

In the present article we also do not analyze cosmological aspects of the 
RS models limiting ourselves only to a few short remarks. 
In a number of papers a Friedman-Robertson-Walker (FRW) generalization 
of the RS1 model was studied (see Refs.~\cite{Cs99}, \cite{RScosm}, 
also Ref.~\cite{AENPSTZ} and an extensive list of references therein). 
Because of the presence of the branes the effective cosmological 
equations on the TeV-brane are non-standard. In particular, the effective 
Hubble parameter $H_{2}$ on this brane turns out to be proportional 
to the square of the energy density $\rho^{2}$ (and not to $\rho$, as in the 
standard cosmology), that for certain regimes leads to a contradiction 
with the standard behaviour of the scale factor of the three-dimensional 
space in the FRW Universe. In addition, the equation for $H_{2}$ 
includes an extra term which can be interpreted as an effective radiation 
term. However, as it was shown in Ref.~\cite{Cs99}, 
if a mechanism, stabilizing the radion (e.g. the Goldberger-Wise 
mechanism, Ref. \cite{GolWi99}), is added, then the standard 
FRW cosmology is recovered for the temperatures 
$T < \Lambda_{\pi}$.

\section{Discussion and Conclusions}

We have considered two classes of models with extra dimensions, the ADD 
model and the RS-type models. The models were designed to solve the hierarchy 
problem. It turns out that they predict some new effects which can be 
detected at the existing (Tevatron) and future (LHC, TESLA) colliders. 
This opens new intriguing possibilities of discovering new physics 
and detecting extra dimensions of the space-time in high energy experiments 
in the near future.  
Their results may give us deeper understanding of the long 
standing hypothesis by T. Kaluza and O. Klein and even provide new 
arguments in favour or against it.    

Limitations on the size of the article have not allowed  
us to include a number of interesting issues related to the 
ADD and RS models. 
In particular, we have not considered the neutrino physics within the 
KK approach. Apparently the idea of neutrinos experiencing extra dimensions 
was first introduced in articles \cite{DDG-ADDM}, and later was developed 
in Refs.~\cite{GrNe99}, \cite{KK-nu}.   
Also topics like the analysis of the anomalous magnetic moment of the muon 
(see Ref. \cite{KK-muon}), latest developments in SUSY extensions  
of the SM with extra dimensions (see Refs.~\cite{PoQui98}, \cite{EWSB}), 
latest results on astrophysics with extra dimensions (see, for example, 
Ref.~\cite{Pos01} and references therein), and many others are left 
beyond the scope of the present review.    

We finish the article with a short summary of main features of the ADD 
and RS models. 

\underline{ADD Model}. 

\begin{enumerate}
\item The ADD model removes the $M_{EW}/M_{Pl}$ hierarchy, but replaces 
it by the hierarchy 
\[
\frac{R^{-1}}{M} \sim \left( \frac{M}{M_{Pl}} \right)^{2/d} \sim 
10^{-\frac{30}{d}}.
\]
For $d=2$ this relation gives $R^{-1}/M \sim 10^{-15}$. 
This hierarchy is of different type and might be easier to 
understand or explain, perhaps with no need for SUSY. 

\item The scheme is viable. 

\item For $M$ small enough high energy physics effects, 
predicted by the model, 
can be discovered at future collider experiments. 

\item For $d=2$ the cosmological and astrophysical bounds on $M$ are 
high enough ($M \geq 100$ TeV), so that a (mild) hierarchy 
is already re-introduced. For $d \geq 3$ the bounds on $M$ are sufficiently 
low. 

\item Some natural cosmological scenarios within the ADD approach may bring 
further problems. One of them is the $m_{inf}/M$ hierarchy, where 
$m_{inf}$ is the mass of the inflaton (see Eq. (\ref{ADD:infl})). 
Another is the moduli problem. These may 
be indications of the need for SUSY in multidimensional theories. 
\end{enumerate} 

\underline{RS1 model}

\begin{enumerate} 
\item The model solves the $M_{EW}/M_{Pl}$ hierarchy problem 
without generating a new hierarchy. 

\item A large part of the allowed range of parameters   
of the RS1 model will be studied in future collider experiments 
which will either discover the RS1 model or exclude 
the most "natural" region of its parameter space (see Sect. 3.2). 

\item With a mechanism of radion stabilization added the model is quite 
viable. In this case cosmological scenarios, based on the RS1 model, 
are consistent without additional fine-tuning of parameters (except 
the cosmological constant problem). 
\end{enumerate} 

\section*{Acknowledgements}

We are grateful to E.E. Boos, V. Di Clemente, S. King, 
C. Pagliarone, K.A. Postnov, V.A. Rubakov, 
M.N. Smolyakov and I.P. Volobuev for useful discussions and
valuable comments. The author thanks the HEP group of the 
University of Southampton, where a part of the review was 
written, for hospitality. The work was supported in part by the
Russian Foundation for Basic Research (grant 00-02-17679)
and the Royal Society Short Term Visitor grant ref. RCM/ExAgr/hostacct.

\end{document}